\documentclass[10pt,preprint]{emulateapj}

\usepackage{ulem}
\usepackage{CJK}

\newcommand{\kms}{\hbox{km s$^{-1}$}}

\shorttitle{A Gaia DR2 Overview of star formation in the Serpens Molecular Clouds}
\shortauthors{Herczeg et al.}

\begin{document}
\begin{CJK*}{UTF8}{gbsn}

\title{An initial overview of the extent and structure of recent star formation within the Serpens Molecular Cloud using Gaia Data Release 2}

\author{Gregory J. Herczeg(沈雷歌)\altaffilmark{1}, Michael A. Kuhn\altaffilmark{2}, Xingyu Zhou\altaffilmark{1,3}, Jennifer Hatchell\altaffilmark{4}, Carlo F. Manara\altaffilmark{5},\\
Doug Johnstone\altaffilmark{6,7}, Michael Dunham\altaffilmark{8}, Anupam Bhardwaj\altaffilmark{1}, Jessy Jose\altaffilmark{9}, Zhen Yuan(袁珍)\altaffilmark{10}}


\altaffiltext{1}{Kavli Institute for Astronomy and Astrophysics, Peking University, Yiheyuan Lu 5, Haidian Qu, 100871 Beijing, People's Republic of China}
\altaffiltext{2}{Department of Astronomy, California Institute of Technology, Pasadena, CA 91125, USA}
\altaffiltext{3}{Department of Astronomy, Peking University, Yiheyuan 5, Haidian Qu, 100871 Beijing, People's Republic of China}
\altaffiltext{4}{Physics and Astronomy, University of Exeter, Stocker Road, Exeter EX4 4QL, UK}
\altaffiltext{5}{European Southern Observatory, Karl-Schwarzschild-Strasse 2, Garching bei München, 85748, Germany}
\altaffiltext{6}{NRC Herzberg Astronomy and Astrophysics, 5071 West Saanich Rd, Victoria, BC, V9E 2E7, Canada}
\altaffiltext{7}{Department of Physics and Astronomy, University of Victoria, Victoria, BC, V8P 1A1, Canada}
\altaffiltext{8}{Department of Physics, State University of New York at Fredonia, Fredonia, NY 14063}
\altaffiltext{9}{Indian Institute of Science Education and Research Tirupati, Rami Reddy Nagar, Karakambadi Road, Mangalam (PO) Tirupati 517507, India}
\altaffiltext{10}{Key Laboratory for Research in Galaxies and Cosmology, Shanghai Astronomical Observatory, Chinese Academy of Sciences, 80 Nandan Road, Shanghai 200030, Peopleʼs Republic of China}


\begin{abstract}
The dense clusters within the Serpens Molecular Cloud are among the most active regions of nearby star formation.  In this paper, we use Gaia~DR2 parallaxes and proper motions to statistically measure $\sim1167$ kinematic members of Serpens, few of which were previously identified, to evaluate the star formation history of the complex.  The optical members of Serpens are concentrated in three distinct groups located at 380--480 pc;
the densest clusters are still highly obscured by optically-thick dust and have few optical members.
The total population of young stars and protostars in Serpens is at least 2000 stars, including past surveys that were most sensitive to protostars and disks, and may be far higher.
Distances to dark clouds measured from deficits in star counts are consistent with the distances to the optical star clusters.
The Serpens Molecular Cloud is seen in the foreground of the Aquila Rift, dark clouds located at 600--700 pc, and behind patchy extinction, here called the Serpens Cirrus, located at $\sim250$~pc.  Based on the lack of a distributed population of older stars, the star formation rate throughout the Serpens Molecular Cloud increased by at least a factor of 20 within the past $\sim5$~Myr.  
The optically bright stars in Serpens Northeast are visible because their natal molecular cloud has been eroded and not because they were flung outwards from a central factory of star formation.
The separation between subclusters of 20--100~pc and the absence of an older population leads to speculation that an external forcing was needed to trigger the active star formation.
\end{abstract}
 

\section{INTRODUCTION}

\begin{figure*}[!t]
\epsscale{1.2}
\plotone{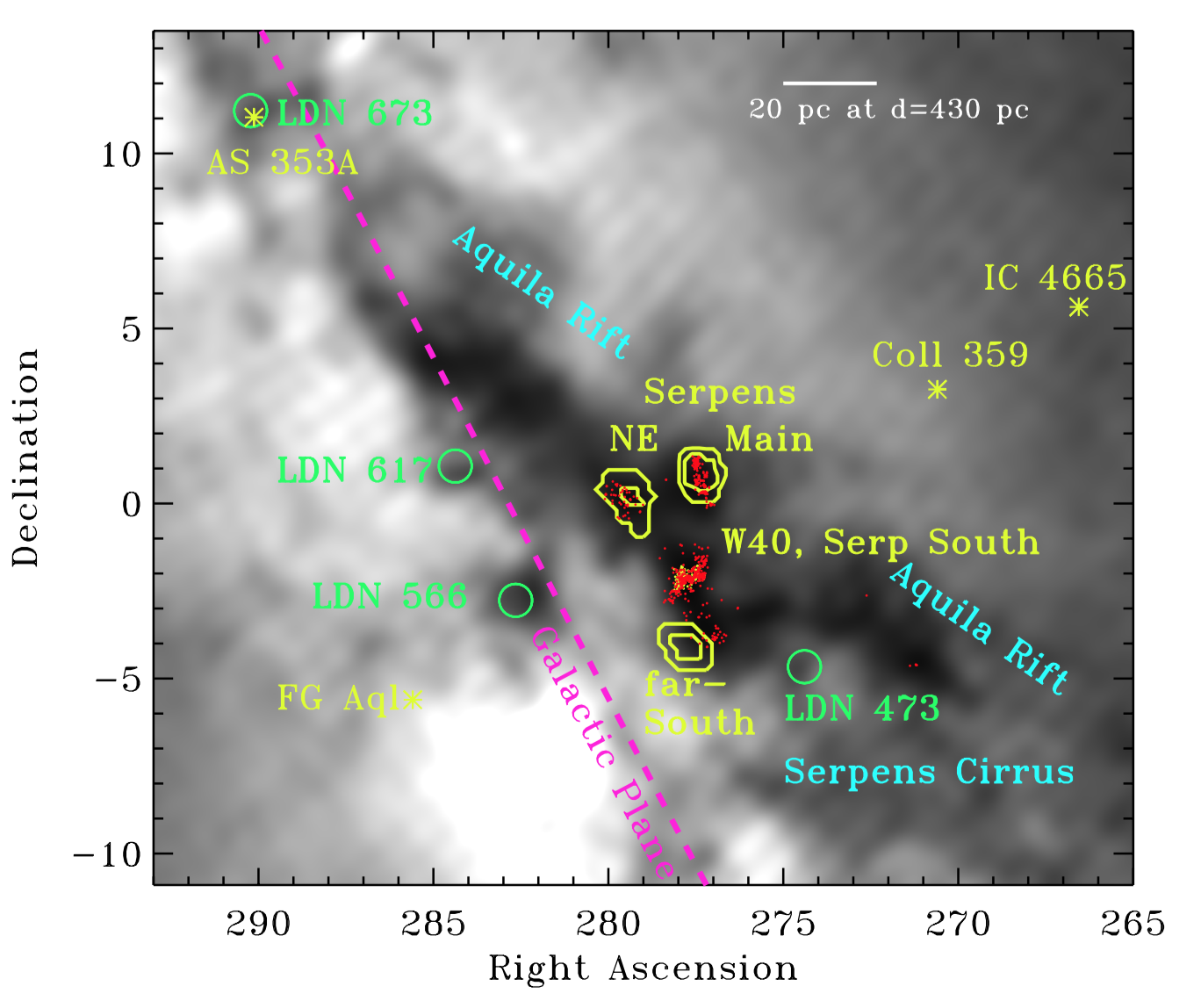}
\caption{A map of the number of stars with distances 1000--1500 pc, behind the Serpens star-forming regions and the Aquila Rift.  The dark regions trace dust extinction, with locations of dust that correlate well with the extinction map from \citet{cambresy99}.  The prominent star-forming regions Serpens Main and Serpens South are labeled on the map.  
Several dark clouds are labeled (green circles), including LDN 673 on the northeast end of the Serpens star-formation complexes (\citealt{lynds62}, see also \citealt{dobashi11}).  The diagonal Aquila Rift cuts across the galactic plane (dashed purple line); the Serpens Cirrus is located in the southwest.   Three clusters of optically-bright stars in Serpens are identified with yellow contours.  The youngest regions, including W40 and Serpens South, are identified on the map by the location of protostars and disks that were identified by \citet{povich13} (yellow circles) and \citet{dunham15} (red circles).}
\label{fig:largemap}
\end{figure*}

The spatial distribution of young star clusters provide constraints for theoretical explanations for the growth of molecular clouds and subsequent star formation \citep[e.g.,][]{vazquez-semadeni18,cunningham18}.
Our descriptions of these star-forming regions are built upon a census of the cloud complexes and stellar population.
The Serpens Molecular Cloud, one of the most active sites of ongoing star formation within 500 pc, is particularly challenging for membership analyses because the region is seen against the crowded background of the galactic plane, is visually located against a long swath of extinction, and suffers from historical uncertainties in the distances to the different clouds.
The {\it Gaia} satellite \citep{gaia2016,gaia2018} is now revolutionizing the description of star-forming regions. Initial results from {\it Gaia} DR2 have described the star formation history of the Orion molecular cloud complexes \citep{kounkel18}, the expansion of individual young clusters  \citep{kuhn18}, of individual sub-clusters in the Vela OB2 Association \citep{cantat19vela}, and of the Gould Belt \citep{dzib18,zari18}, the distinction between sub-groups within single regions \citep{beccari18,veronica18}, and memberships of specific regions for analysis of disk fractions \citep{manara18} and older populations \citep{damiani18,luhman18}.

In this paper, we apply the power of {\it Gaia} astrometry to star formation in the Serpens-Aquila complex.  This collection of cloud complexes\footnote{We adopt the following nomenclature for this region:  Serpens is the collection of star-forming regions and optically-thick molecular clouds at 370--500 pc, located roughly in a rectangle from 276--281 deg in right ascension and from -5 to +2 deg in declination. The Aquila Rift is the elongated swath of extinction that cuts diagonally across most of our field and is located at 600--700 pc.  Serpens Cirrus is a more diffuse cloud to the southwest of Serpens and is located at $\sim 250$ pc.} 
can be subdivided into several distinct clouds and star-forming regions (see the map of Gaia DR2 star counts in Figure \ref{fig:largemap} and extinction maps in \citealt{cambresy99} and \citealt{green15}). The Serpens Main star-forming region, located on the northern side of the rift, has historically been the most well-studied site of active star formation within the complex \citep[see review by][]{eiroa08}.  
About 3$^\circ$ to the south of Serpens Main, a separate cloud contains the \ion{H}{2} region W40 on its east side and Serpens South $\sim 30^\prime$ to the west.  W40, discovered in surveys of 22 cm radio continuum emission and H$\alpha$ emission \citep{westerhout1958,sharpless1959}, includes a compact cluster of $\sim$500 candidate young stars \citep{kuhn10,mallick13} and is ionized by several OB stars \citep{smith85,shuping12}.  Serpens South, a dense embedded cluster in a massive molecular filament, was discovered in the {\it Spitzer} survey of \citet{gutermuth08ser} and is rich in protostars \citep[see also][]{dunham15,konyves15}.  Some young stars and \ion{H}{2} regions are also located in a region around MWC 297 \citep[e.g.][]{bontemps10,dunham15,rumble15}, hereafter called Serpens far-South. A fourth group of the cloud complex, which we designate Serpens Northeast based on its location 1$^\circ$ east of Serpens Main, was found by \citet{dunham15} to have its own population of young stars.  These groups of young stars are visually interspersed within the Aquila Rift, a set of dark molecular clouds elongated across $>25^\circ$ on the sky (260 pc at a distance of 600 pc).

The physical connection between these complexes has been questioned, in part because the distances to Serpens Main, Serpens South and W40, and the Aquila Rift have been controversial (see discussions in \citealt{eiroa08}, \citealt{prato08}, and \citealt{RodneyReipurth08}).  
The VLBI parallax measurements of masers by \citet{ortiz17ser} lead to (separate) distances of 436 pc to Serpens Main and to W40.
The similarity in velocities indicate that the entire region is located at the same distance \citep[see discussion in][]{ortiz17ser}.
However, a recent analysis of the X-ray luminosity function of Serpens South placed the cluster at 260 pc \citep{winston18}, consistent with the previous identificiation of an extinction wall at $\sim 250$ pc to both Serpens Main and the Aquila Rift \citep{straizys96,straizys03}.  In the 3D structure of the Milky Way from the velocity analysis of \citet{lallement14}, the Aquila Rift is also placed at $\sim 250$ pc.  The possible connections in the wider complex have also suffered from uneven accounting of stellar membership; many dedicated surveys on smaller regions 
\citep[e.g.][]{kuhn10,povich13,mallick13,dunham15} miss most optical members of Serpens Northeast and Serpens far-South that are found in this paper.

With all-sky coverage, {\it Gaia} DR2 provides a consistent and near-complete, magnitude-limited set of astrometry and photometry over a full field of interest.
When compared to past membership studies, our application of {\it Gaia} DR2 astrometry is particularly sensitive to the lightly-absorbed population, including the disk-free (Class III) young stars that are missed in searches based on excess infrared emission.  Our search over a wide field of view probes both recent star formation and any population of older stars from in past epochs of star formation.
In this paper, we use {\it Gaia} DR2 almost {\it exclusively} to identify and analyze the population of young stars in the Serpens star-forming regions, and the lack of young stars in the Serpens Cirrus.  Distances to different subclusters and to dust extinction regions confirm the accurate VLBI distance measurements of \cite{ortiz17ser} to the Serpens star-forming regions\footnote{A contemporaneous and independent paper by \citet{ortiz18ser} uses Gaia DR2 data to confirm the VLBI distances to Serpens Main, Serpens South, and W40, consistent with our results.} and reveal the presence of some dust at 250 pc that is not actively forming stars.  Most of the optically-thick dust outside of the active Serpens star-forming regions are located at $\sim 600-700$ pc.

\begin{figure}[!t]
\epsscale{1.25}
\begin{center}
 \includegraphics[width=0.55\textwidth, trim=85 90 20 280,clip]{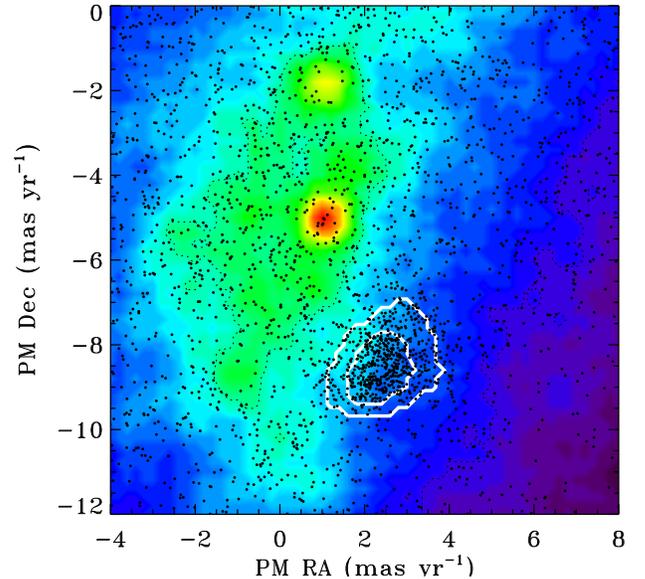}
\end{center}
\vspace{-1mm}
\caption{A heat map of the proper motions of all stars in the full downloaded area with distances
  between 350--500 pc and $\varpi/\sigma(\varpi)>10$.  The proper motion of stars at this distance
  around the Serpens star-forming region (right ascension of 276-281 and declination from -5 to 2 deg.; solid white contours; black dots show individual stars) are
  offset in proper motion from this field population.  For both the Serpens and field samples, contours are shown for 0.2 and 0.5 times the peak number of stars in a $0.2\times0.2$ (mas/yr)$^2$ box.  The two bright overdensities in the heat map are the older clusters IC 4796 and NGC 6633.}
\label{fig:propermotions}
\end{figure}

\section{Sample and Data}

Our analysis in this paper is based primarily on data from {\it Gaia} DR2.  We downloaded all stars in the {\it Gaia} DR2 archive \citep{arenou18} in a large area around
the star-formation complexes.  We began by downloading 1 square degree
around both Serpens Main and W40/Serpens South clusters.  Our desire
to describe the dust extinction in the context of the broader cloud led us to download the surrounding few square degrees.  Several iterations later, we had downloaded all $\sim 140$ million objects in $\sim 830$ square degrees around Serpens-Aquila, from $\alpha=262$ to $294$ degrees in right ascension and $\delta=-12$ to $14$ degrees in declination.  
Star count maps demonstrate that the Aquila Rift extends beyond this region, but we had to stop for our own sanity and so that IDL could run with data arrays of the entire population.

\begin{figure*}[!t]
\epsscale{1.25}
 \includegraphics[width=0.49\textwidth, trim=40 370 40 40,clip]{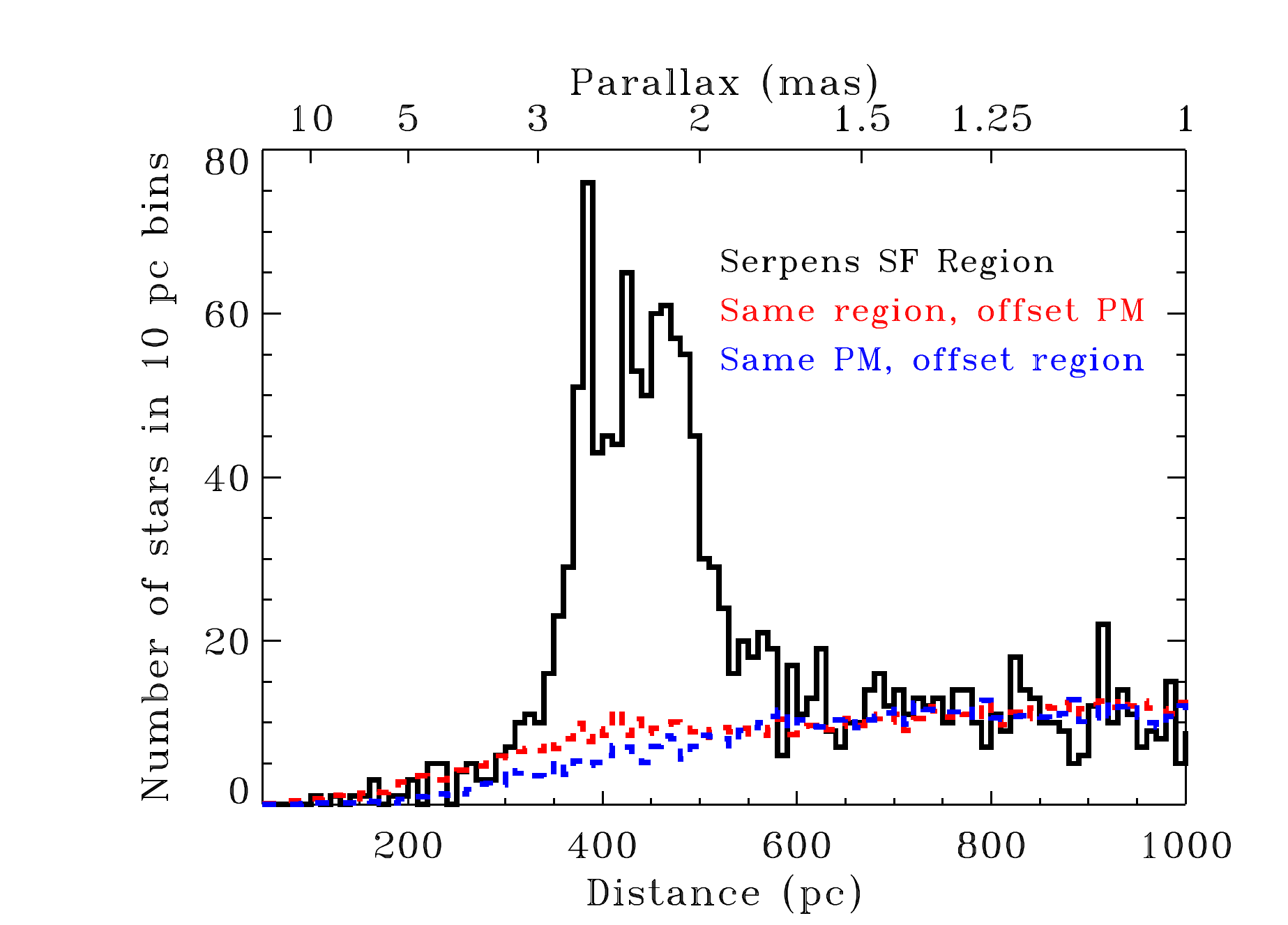}
 \includegraphics[width=0.49\textwidth, trim=40 370 40 40,clip]{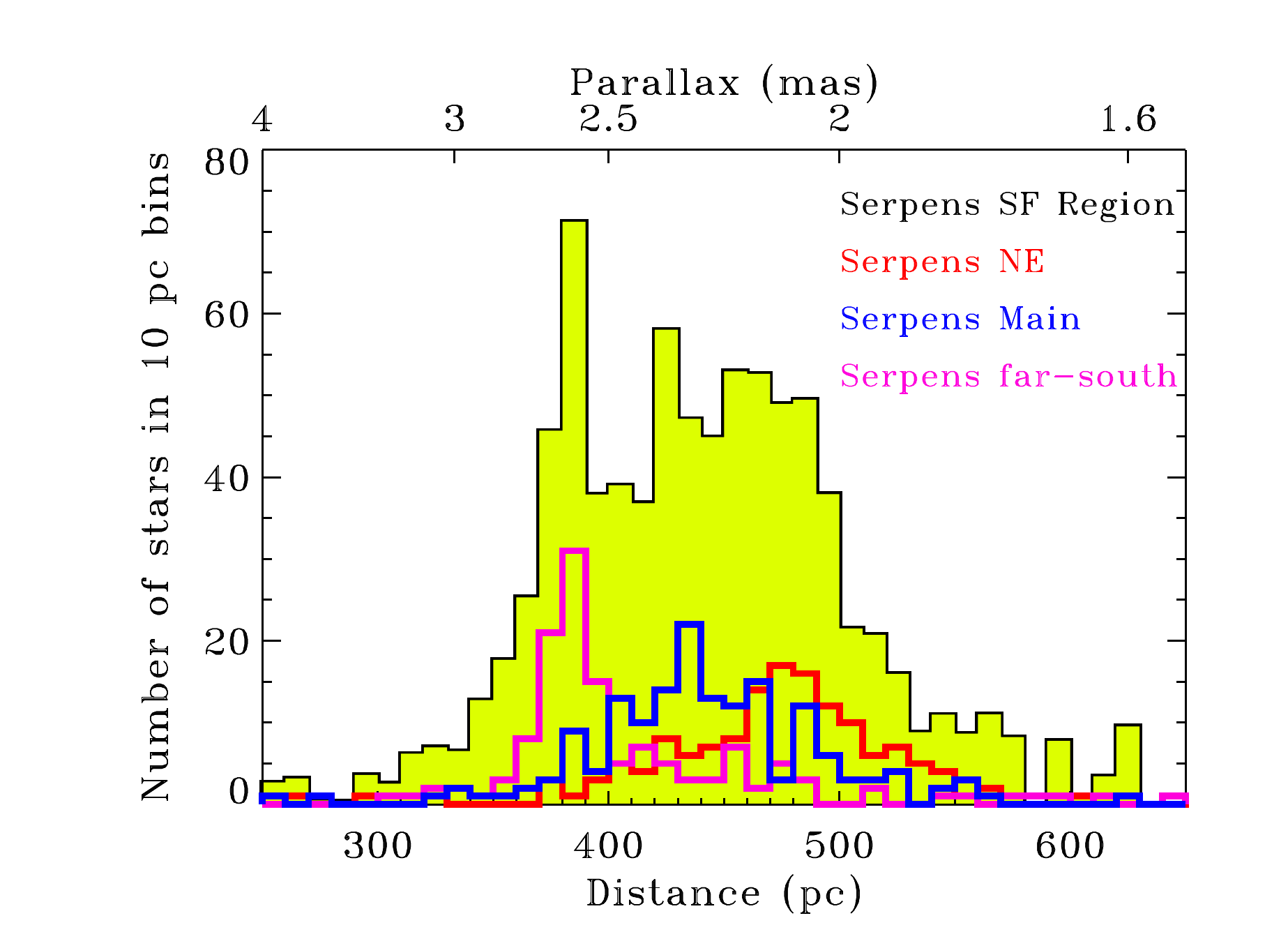}
\caption{{\it Left:}  Star counts versus distance for stars with proper motions within 2 mas/yr of the Serpens star formation centroid (black solid line, see centroid in Fig.~\ref{fig:propermotions}), compared to star counts for the full downloaded area outside of the Serpens star-forming regions (blue dashed line) and to star counts from the Serpens star-forming region but with offset proper motions.  
 The off-cloud and offset proper motion populations are scaled to that of the Serpens star-forming region at 150--300 and 700--900 pc.  The excess number of stars from 350--500 are the young members.  {\it Right:}  Three
  peaks in the histogram of star counts versus distance correspond to three different sub-clusters: Serpens Main, Serpens Northeast, and Serpens far-south (each restricted to within 1 mas/yr in proper motion and 1 degree of the centroid of each subregion, see also Table~\ref{tab:clusters} and Figure \ref{fig:serpstars}.)}
\label{fig:distclusters}
\end{figure*}

The most relevant data are the stellar positions, proper motions,
parallaxes, and $G$, $Bp$, and $Rp$-band magnitudes, along with the associated errors and the excess astrometric noise \citep{lindegren18,evans18gaia}.  The optical photometry from {\it Gaia} is supplemented with 2MASS $JHK_S$ \citep{cutri03}.  Objects
fainter than $G=19$ mag are not considered in the bulk analysis because
the unreliability of small parallaxes of faint objects leads to an
artificially large population of stars at $<300$ pc, even after
accounting for the listed errors.\footnote{
The star-count map for all stars with $G<19$ does not show any evidence
for optical extinction until $\sim 250$--$400$ pc to the Serpens star-formation complex; however, for fainter stars, the extinction regions are prominent in density maps of stars with 
parallaxes $>10$ mas ($<100$ pc).  The tail of the error distribution in parallaxes leads to a background population that would be interpreted as nearby objects.}  
Most analyses in this paper are also restricted to stars with 
excess astrometric noise less than 2 mas
and to stars with parallaxes measured to a signal-to-noise greater than
5--20, depending on the analysis.  However, fainter stars and stars with parallaxes consistent with 0
are used when evaluating membership of candidates selected from past surveys.  The systematic error of $\sim0.1$ mas in {\it Gaia} DR2 parallax measurements \citep[e.g.][]{luri18,leung19} is ignored in our analysis and is applied only in the final measurements of distance. 

All star count maps created from {\it Gaia} observations in this region include diagonal dips in the number of stars brighter than $G=19$.  These features must be related to some loss in sensitivity or astrometric accuracy, presumably at the edges of the {\it Gaia} field of view, as {\it Gaia} scans across the sky.  The depth and accuracy also depends on star density, which varies strongly across our field.  We ignore these systematic problems. 

For ease of reference, at 430 pc a distance of one degree corresponds to 7.5 pc, while a proper motion of 1 mas/yr corresponds to 2 km/s.

\section{A search for young stars in Serpens}

In this section, we identify several young clusters within our field in a search
of the 5-D phase space of location, 
proper motion, and distance.  
The ability to separate the young stars in the Serpens in both
distance and proper motion provides us with the leverage to distinguish
the young star population from the field population, without resorting
to spectroscopic or even photometric criteria \citep[see, e.g. seminal papers by][]{blaauw56,dezeeuw99}. Photometry is not directly used in these selections, beyond the initial {\it Gaia} selection of targets and our brightness requirement of $G<19$ mag.  However, the stars in these clusters are located above the main sequence in color-magnitude diagrams (see Section 5.2), which reinforces the conclusion that they are young.

Although the density of stars is high towards Serpens, the members of the star-forming regions and young clusters are readily
identified in both proper motion and distance.  Figure~\ref{fig:propermotions} shows the proper motions of all stars in the full field 
with distances $\sim 350-500$ pc (see Fig.~\ref{fig:distclusters} for this circular logic) and parallaxes measured to better than $\varpi/\sigma(\varpi)>10$, for the full field (except Serpens) and for a box around Serpens,
between $\alpha=276$ to $281$ degrees in right ascension and $\delta=-5$ to $2$ degrees in declination.  A clear over-density of stars is visible at PM(RA)$\approx2.5$ mas/yr and PM(Dec)$\approx-8.5$ mas/yr.  In this initial analysis, this Serpens box includes 470 stars with proper motions within 1 mas/yr of the Serpens centroid; the same distribution as the rest of the field would have yielded only $36$ stars in the same region.  The distribution of parallaxes of stars within this box (Figure~\ref{fig:distclusters}) shows a large excess at 350--500 pc, the expected distance for the Serpens star-forming region.

\begin{figure*}[!t]
\begin{center}
\epsscale{0.5}
 \includegraphics[width=0.7\textwidth]{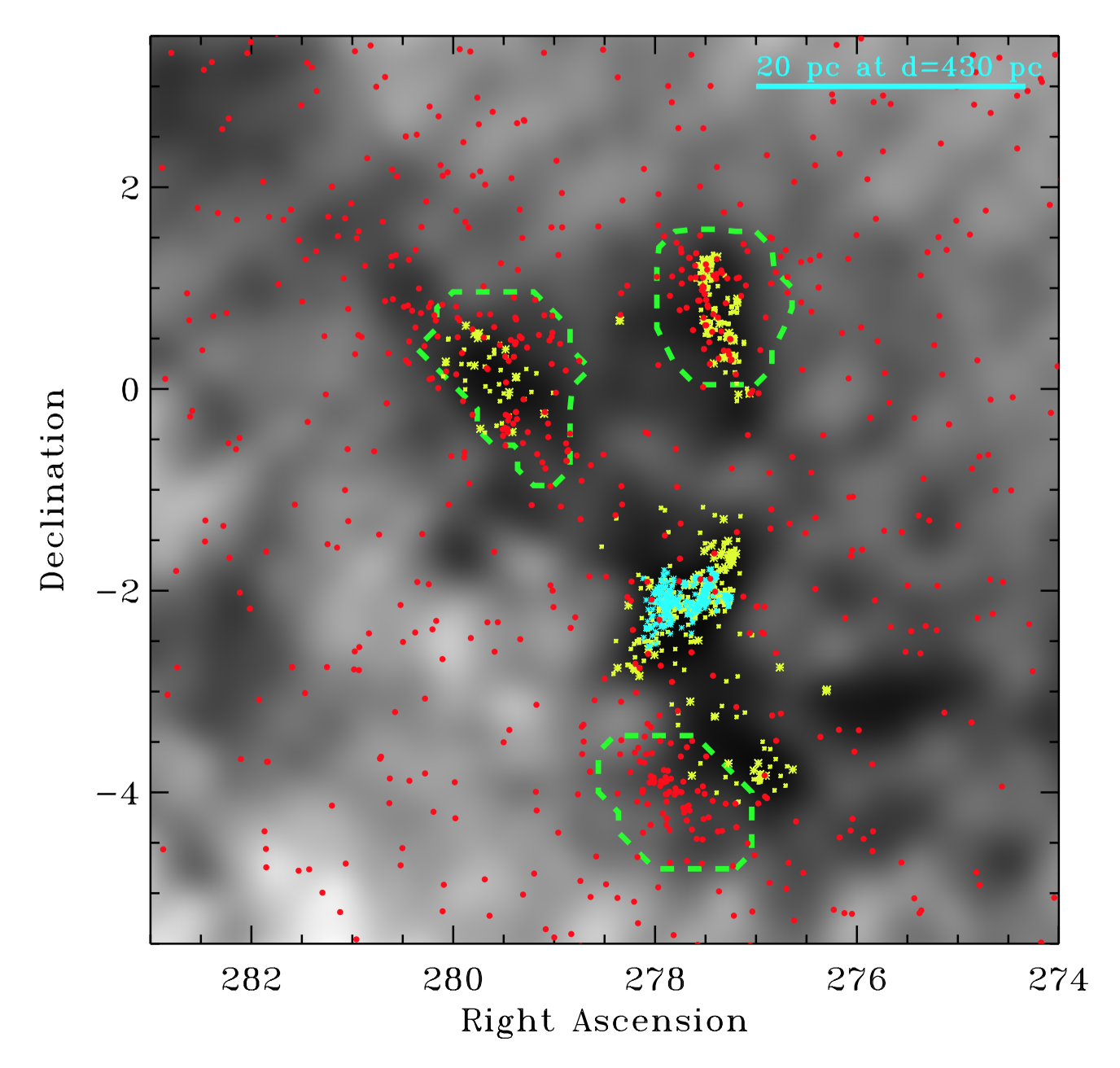}
\end{center}
\vspace{-5mm}
\caption{The spatial location of stars with $\varpi/\sigma(\varpi)>10$, distances between 350--550 pc, and proper motions within 2 mas/yr of the centroid proper motion of the Serpens star-forming region (red circles), compared with the distribution of Class 0/I/II protostars identified from mid-IR photometry (small yellow asterisks as disks and large yellow asterisks as protostars from \citealt{dunham15}, and blue asterisks from \citealt{povich13}).  The optical members are located in three distinct clusters:  Serpens Main, Serpens Northeast, and Serpens far-South (outlined by the dashed green contours established from stellar densities).
Many protostars but few optical counterparts are associated with the optically-thick Serpens South and W40 molecular clouds, perhaps indicating that these regions are younger than the other clusters.  Field stars are contaminants and are spread roughly equally throughout this region.}
\label{fig:serpstars}
\end{figure*}

This overdensity in proper motion and parallax space corresponds to the optical members of the Serpens star-forming region.  Figure~\ref{fig:serpstars} shows the spatial distribution of stars between 350-550 pc and with a
proper motion within 2 mas/yr of the centroid proper motion.  Three distinct regions stand out: the optical counterparts of Serpens Main and Serpens Northeast, as well as a region associated with MWC 297 and Sh 2-62, here called Serpens far-South (also sometimes called the MWC 297 group or region, \citealt{rumble15}).  The intervening drop in stellar density between Serpens Main and Serpens Northeast does not correspond to any increase in extinction, which demonstrates that the two subclusters are distinct.

The properties of each sub-cluster are listed in Table~\ref{tab:clusters}.  The distance, location, and proper motion are first
estimated with loose distance and proper motion criteria for a
sample of stars with parallax measured to $\varpi/\sigma(\varpi)>20$.  The distance is then refined from
parallaxes of a restricted candidate membership criteria, restricted
to stars with $2\sigma$ (here $\sigma$ is the scatter in proper motions, not an uncertainty) of the proper motion centroid and within 
$0.5$--$0.8$ degrees of the spatial centroid of the cluster.  The proper motion is 
refined with a similar membership criteria, with a $2\sigma$ restriction in distance rather than proper motion.  The final proper motion median and standard deviation is calculated by fitting a Gaussian profile to the proper motion in both right ascension and declination.  The final distance measurement is calculated by a weighted average of the parallax, with a statistical uncertainty of 1--2 pc.  The listed standard deviation in the distance is the standard deviation in a distance spread that is required to reproduce the observed spread in parallaxes.  The distance spread for Serpens far-South is 2 pc, indicating that the measured distance spread is nearly consistent with the spread expected from only the parallax errors of the members.  Other regions have distance spreads that are larger than the projected radius of the cluster, which indicates that either the parallax errors are underestimated, that a population of distributed young stars within Serpens contaminates the sub-cluster population, or that the sub-clusters are elongated along our line of sight.

\subsection{The Total Optically Bright Population of the Serpens Star-forming Regions}

\begin{figure}[!t]
\epsscale{1.3}
\includegraphics[width=0.49\textwidth, trim=40 370 40 40,clip]{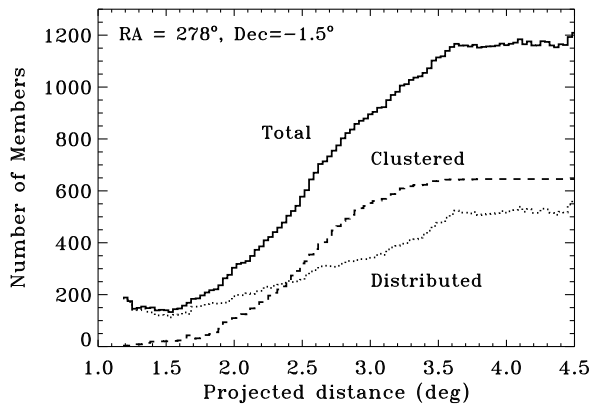}
\caption{The cumulative radial distribution of members from the approximate projected centroid of Serpens star formation at right ascension $\alpha=278^\circ$ and declination $\delta=-1.5^\circ$.  The number of members are split into a total clustered population of stars in Serpens Main, Serpens Northeast, and Serpens far-South, and a distributed population of stars that are not affiliated with those three optical clusters.  The distributed population includes a few likely members of Serpens South and W40.  Any epochs of vigorous star formation in the past $\sim 30-50$ Myr would have been identified either as a cluster or as a distributed population of stars.  These memberships are statistical and therefore account naturally for interlopers.  Some small numbers of stars are even more extended than shown here, including the LDN 673 and FG Aql groups.}
\label{fig:radial}
\end{figure}

In this subsection, we estimate the total number of stars in our sample that are located within each of the sub-clusters identified above, as well as the total number of stars within the Serpens Star-forming region.
These populations measured by estimating a radial size of the cluster using star counts and subsequently measuring the number of stars from the distribution of proper motions within a circular area (Table~\ref{tab:clustersize}).
The stars are best separated from field stars in the proper motion in declination, so the total number of stars is measured by fitting a Gaussian profile to that distribution within some circular area on the sky.  Figure~\ref{fig:radial} shows that the stars extend to $\sim 3.2^\circ$ (26 pc) for 90\% completeness, from an approximate center of $\alpha=278^\circ$ and $\delta=-1.5^\circ$.  No significant population of young stars is detected between 3.5--4.5$^\circ$ of this location.  A small population of young stars in Serpens is located beyond $4.5^\circ$ and is discussed in \S 3.2.

The Serpens star-forming region includes 1167 members (evaluated statistically) with {\it Gaia} DR2 parallaxes with $\varpi/\sigma(\varpi)>5$.  Roughly 60\% of those stars are located within one of the three main optical clusters; a few additional stars are likely members of the W40/Serpens South cluster.  The remaining stars are distributed throughout the region.  
The accounting of optical members likely misses at least half of the full membership, for many reasons.  The color-magnitude diagrams (discussed in \S 5) indicate a sensitivity to stars with masses larger than 0.2--0.3 M$_\odot$. The embedded protostars are optically-faint and not detectable with {\it Gaia} (see \S 6.1).  Our accounting also misses almost all members of W40 and Serpens South \citep{povich13,mallick13,dunham15}, including any diskless population, because they are buried deep in the molecular cloud.  Any stars on the back side of dark clouds would also be too faint for {\it Gaia}.

\begin{table*}
\begin{center}
\caption{Properties of Stellar Clusters}
\label{tab:clusters}
\begin{tabular}{lccccccccc}
Name & $R_{pars}$ & RA & Dec & PM(RA) & $\sigma$(PM RA)$^b$ & PM(Dec) & $\sigma$(PM Dec)$^b$ & $d$ & $\sigma(d)$$^c$ \\ 
& deg & deg & deg & mas/yr & mas/yr & mas/yr & mas/yr & pc & pc\\
\hline
Serpens Main &  0.5 & 277.51 & 1.08 & 3.08 & 0.57 & -8.55 & 0.61 & 438 & 11\\
Serpens Northeast  & 0.8 & 279.54 & 0.37 & 2.61 & 0.41 & -8.05 & 0.40 & 478 & 6 \\
Serpens far-South & 0.7 & 277.91 & -3.88 & 2.05 & 0.25 & -8.96 & 0.28 &  383 & 2\\
LDN 673$^a$ & 0.5 & 290.25 & 11.28 & 2.46 & 0.32 & -10.25 & 0.24 & 407 & 16\\
\hline
\multicolumn{10}{l}{$^a$: values calculated for all sources with parallaxes of $S/N>10$.}\\
\multicolumn{10}{l}{$^b$$\sigma$ for proper motions are the standard deviation of the population.}\\
\multicolumn{10}{l}{$^c$$\sigma$ for distance is the standard deviation of the population, after correcting for uncertainties in parallax.}\\
\multicolumn{10}{l}{~~~The systematic error in parallax of 0.1 mas dominates the uncertainty in the distance.}\\
\end{tabular}
\end{center}
\end{table*}

\begin{table}
\begin{center}
\caption{Populations of Stellar Clusters}
\label{tab:clustersize}
\begin{tabular}{lccc}
Name & $N_{20}^a$ & $N_{5}^a$ & $R_{stars}$ (deg)\\ 
\hline
Serpens Main & 81 & 265 & 1.2 \\
Serpens Northeast  & 80  & 251  & 1.5 \\
Serpens far-South & 73 & 132 & 1.3  \\
Distributed & 105 & 519 & -- \\
\hline
Total Serpens Population$^b$    & 339 & 1167 & 3.6\\
\hline
\multicolumn{4}{l}{$^aN_{20}$ and $N_5$: number of members with $\varpi/\sigma\varpi>20$ and $>5$.}\\
\multicolumn{4}{l}{~~~Uncertainties $\sim 5\%$, see text}\\
\multicolumn{4}{l}{$^b$ Includes the three sub-clusters and the distributed population}\\
\end{tabular}
\end{center}
\end{table}

While the statistical measurement on the number of members is
accurate, the membership probability for any individual star depends
on the field population.  This probability is higher for stars with 
parallaxes measured with $\varpi/\sigma(\varpi)>20$ and for those with proper motions and locations near the respective centroids of each cluster.  The total population from this specific analysis is accurate to $\sim 5\%$, with the uncertainty dominated by the uncertainty in defining the region in physical, proper motion, and distance space.  The subtraction of field stars does not significantly contribute to these errors.   More advanced techniques will eventually assess membership probabilities and then analyze cluster properties based on those probabilities, but the use of python codes is beyond the scope of the career of the first author.

This approach to identifying cluster members excludes any runaway stars that were ejected with proper motions larger than 2 mas/yr (4 km/s) away from the centroid proper motion of the cluster.  While such stars are interesting for understanding cluster dynamics, they comprise a negligible fraction of the total population of a young cluster.  The distribution of proper motions for these young clusters yield a 90\% completeness within 4 \kms.
Our analysis also preferentially excludes many stars with disks that are viewed edge-on.  Variable young stars with associated nebulosity also have unreliable astrometry (e.g., V883 Ori, Lee et al.~2018; Sz 102, \citealt{fang18}), and would therefore tend to have high astrometric errors and/or proper motions and parallaxes that suggest non-membership.  Many binaries will also have been excluded from the {\it Gaia} DR2 astrometric catalog.

\begin{figure*}[!t]
\vspace{-7mm}
\epsscale{1.15}
\plottwo{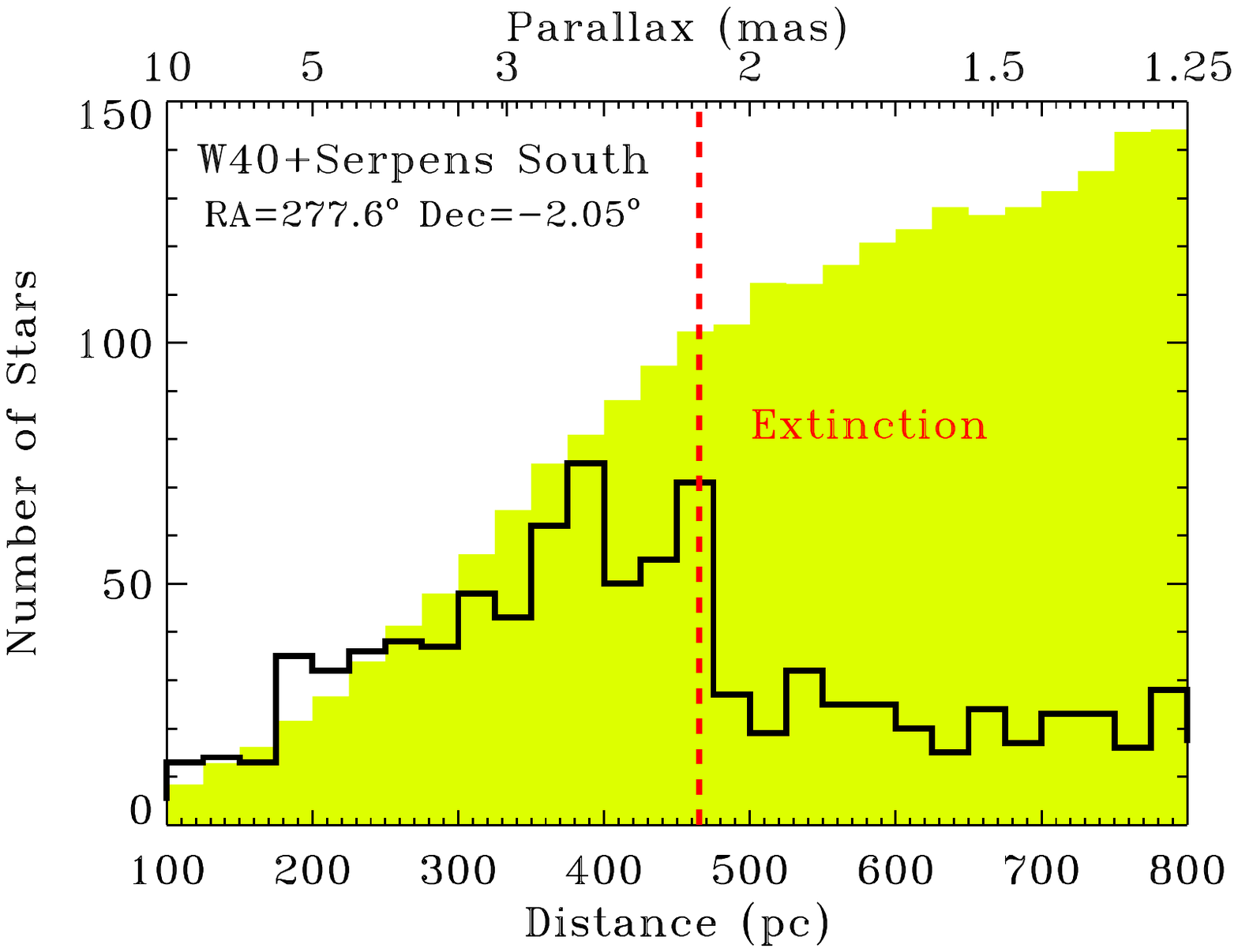}{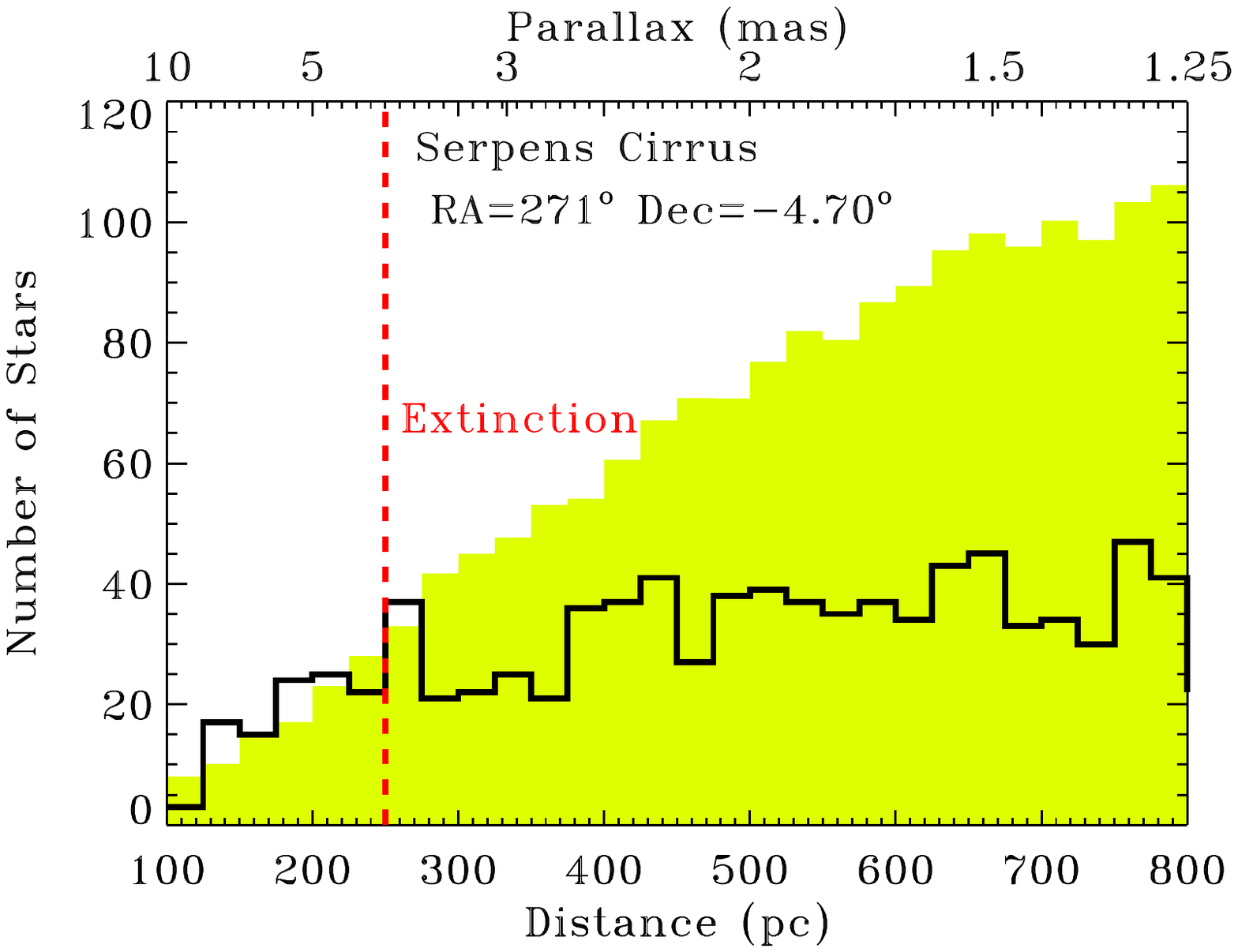}
\vspace{-50mm}
\caption{The number of stars versus distance to (left) Serpens South and (right) Serpens Cirrus (black histograms), compared to off-cloud regions (yellow shaded histogram) that are scaled to the number of stars in Serpens South and Aquila between 50--250 pc, for stars with $\varpi/\sigma(\varpi)>5$.  The star counts versus distance toward Serpens South drops sharply at $\sim460$ pc because of the dark cloud; the apparent, though modest discrepancy in star counts toward Serpens South at $\sim 250$ pc is artificial and well explained by synthesizing populations with a dark cloud at $\sim 460$ pc (see Figure \ref{fig:contourdist} in the Appendix).  The star counts in Serpens Cirrus fall off around 250 pc.  Note that in Appendix B, the region of Serpens Cirrus is a different location, and Figure~\ref{fig:contourdist} shows results for stars with $\varpi/\sigma(\varpi)>20$.}
\label{fig:starctsextinction}
\end{figure*}

\subsection{Searching for isolated groups of stars}

In the previous subsections we identify subclusters and distributed populations of young stars in the Serpens Molecular Cloud.  Past epochs of star formation may have produced populations that have since traveled away from the current epoch of star formation.  The absence of these older populations places strong constraints on the star formation history, discussed in \S 5.4.

Some small groups of stars that are distant from the general boundaries of the active Serpens star-forming region may also be related to Serpens.  A small overdensity of stars related to Serpens Northeast extends $2$ deg (16 pc) northeast of the main cluster, while the Serpens far-South overdensity extends $\sim 3$ deg to the southwest.  A group of $\sim 15$ stars around LDN 673, including AS 353Aab, and a group of 5 stars around FG Aql have proper motions and distances consistent with Serpens membership, but are located $\sim 100$ pc away on the sky.  These small groups are identified here only because of prior knowledge of their likely youth \citep{rice06}, and they would otherwise be challenging to detect.  Other small groups of 5--20 young stars are likely to be present throughout the Serpens-Aquila field. 

Expanding the proper motion cut to 5--10 mas/yr within the centroid of Serpens leads to the identification of IC 4665, IC 4756, and NGC 6633 \citep[e.g.][]{dias02}. The proper motions and ages of these clusters\footnote{The ages of \citet{dias02} should be accurate, since their pre-{\it Gaia} distances are within 5\% of their {\it Gaia} DR2 distances.} together indicate that they are not directly associated with the clouds that are now forming stars in Serpens.  Searching for clusters from $500$ to $600$ pc yields Collinder 359 \citep[e.g.][]{lodieu06,cantat18}, a young cluster located $\sim551$ pc that may be related a previous epoch of star formation in Serpens.

No significant sub-clusters of young stars are identified in Serpens Cirrus, located at $\sim 250$ pc (see \S 4). Our blind search may have missed clusters at proper motions more than 5 mas/yr away from the centroid of the Serpens proper motion.  However, the stars between 225--275 pc ($\varpi/\sigma(\varpi)>20$) are distributed smoothly across the downloaded area.  When divided into $0.5\times0.5$ square degree regions, the stellar distribution is nearly consistent with a Poisson distribution with 6.5 stars per 0.25 square degree.  The proper motion distribution also does not show any significant excess.  A small number of young stars may be present and distributed across the region, however, no clusters are present with $>15$ young stars.

Our search for young stars does not include a rigorous evaluation of populations at $>500$ pc, including whether any young stars or star formation is associated with the Aquila Rift ($d=600-700$ pc).

\section{Measuring the distance to dust extinction}

In the previous section, we identified likely members of the Serpens-Aquila star-forming region, located at 350--500 pc and sharing a common proper motion.  Dust extinction occurs over much wider areas than the stellar clusters.  Clouds may be located at different distances than the visually-associated stellar clusters, and in any star formation region with a distance spread, the optical 
counterparts of a molecular cloud will be biased to members that are in the foreground of the cloud.  In this section, we exploit the stellar distances of field stars to quantify the location and distance to dust extinction in the Serpens and Aquila clouds.

Figure \ref{fig:starctsextinction} shows histograms of star counts versus distance for stars within a region, compared to an off-cloud from within $0.5^\circ$ in galactic latitude located south of the galactic plane.  The star counts include stars with proper motions more than 3 mas/yr (6 km/s at 400 pc) from the centroid proper motion of Serpens members and with parallaxes with $\varpi/\sigma(\varpi)>20$.
The proper motion cut ensures that our selection focuses on field stars and is not influenced by any increase in young stars that are associated with the Serpens-Aquila complex.  For stars within $0.7$ degrees of Serpens South, the star counts drop dramatically at $\sim 460$ pc, indicating that the dust extinction occurs at approximately that distance.  On the other hand, southwest of Serpens, the star counts deviate from the off-cloud region at $\sim 250$ pc.  

This comparison establishes that clouds are present at multiple distances, and that measuring cloud distances with star counts is feasible.  In this section, we first map the star counts with different distance cuts to determine the extent of the Aquila and Serpens clouds.  We then quantify the distance to the dust extinction by simulating populations in different lines-of-sight.

\begin{figure*}[!t]
\epsscale{1.1}
\plottwo{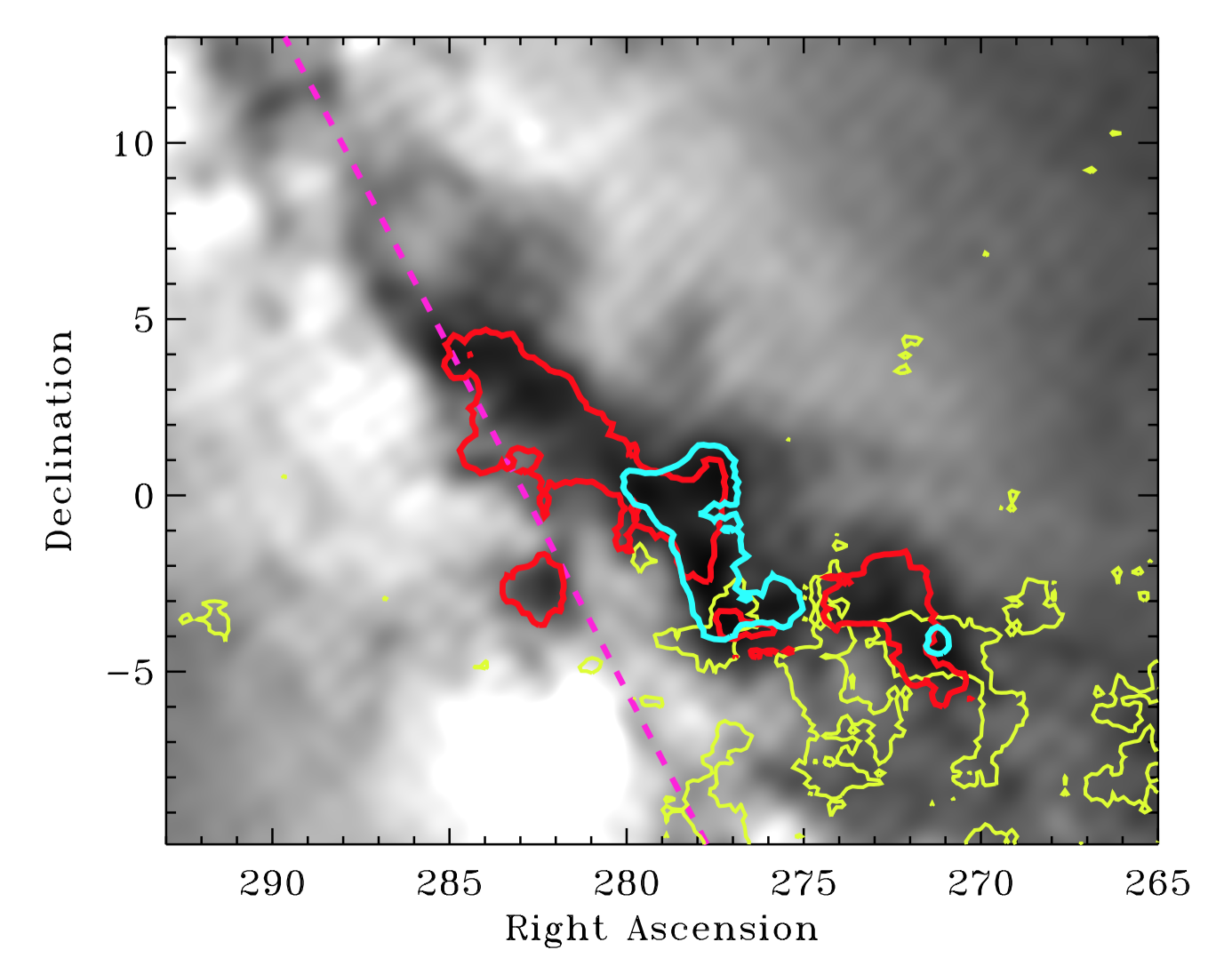}{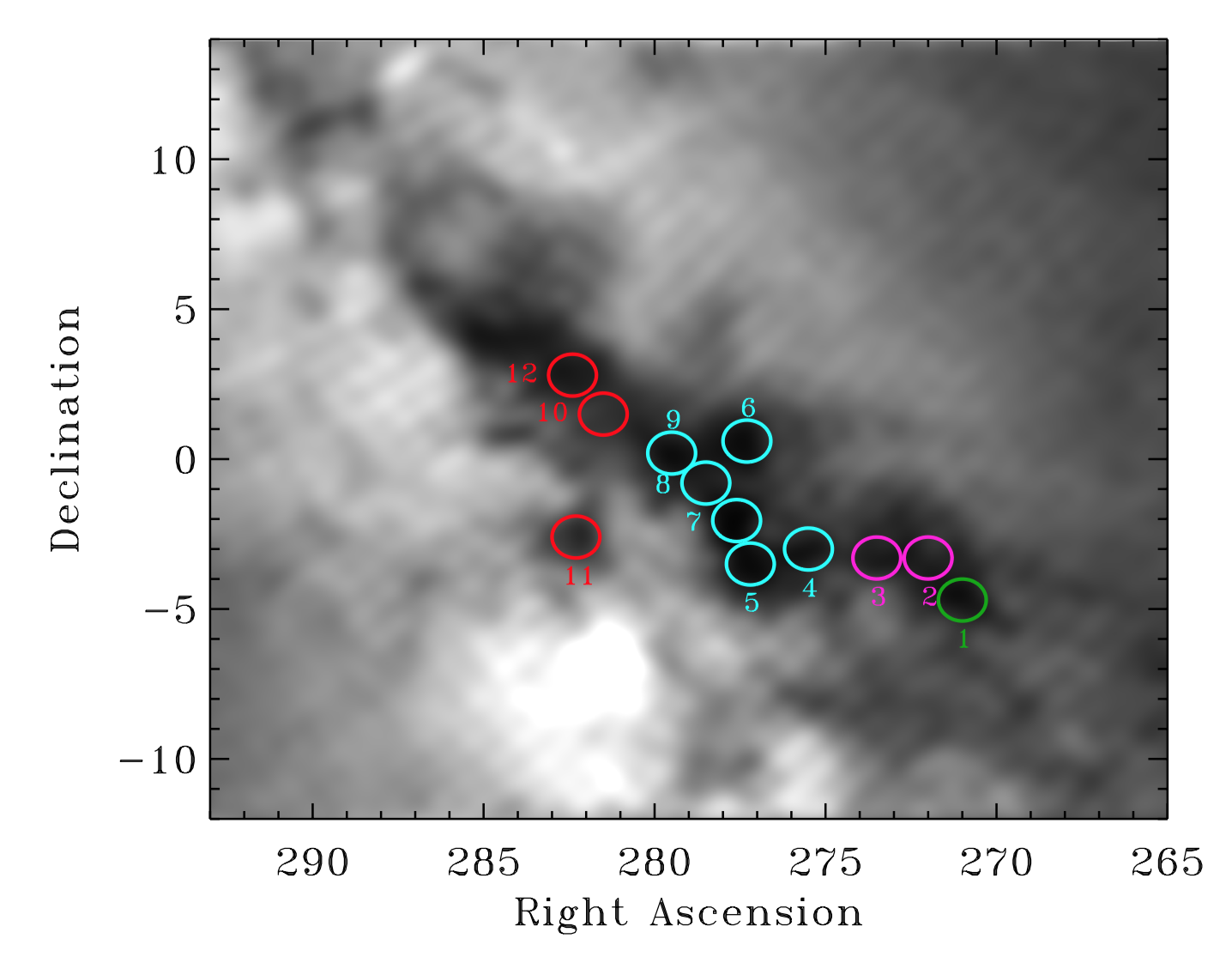}
\caption{{\it Left:} Contours of star-count ratios, $N(d_2-d_3))/N(d_0-d_1))$, as a proxy for extinction at different distances.  All values are corrected for the expected value at each galactic latitude.  The yellow contour shows 
$N(300-400 {\rm pc})/N(30--250{\rm pc})=0.7$ 
for extinction at $\sim 250$ pc, by comparing the number of stars 
between 300--400 pc to those between 30--250 pc.  The blue and red contours show $N(500-600 {\rm pc})/N(300-400 {\rm pc})=0.4$ and 
$N(700-800 {\rm pc})/N(500-600 {\rm pc})=0.4$, 
both more optically thick than the yellow contours.  The contours are created with $0.2\times0.2^\circ$ bins and smoothed with a 2-dimensional Gaussian profile with a full-width half-max of 9 bins.  {\it Right:}  The location of clouds with distances measured and listed in Table~\ref{tab:dustdist}, color-coded with green ($d<300$ pc), cyan (350--480 pc), pink (480--550 pc) and red ($>550$ pc).}
\label{fig:extcontours}
\end{figure*}

\subsection{Three-dimensional maps of dust}

Figure~\ref{fig:extcontours} shows contours of the ratio of star counts with a range of distances, as a proxy for extinction at $\sim 275$ pc, $\sim 450$ pc, and $\sim 650$ pc.  At each location in the map, the total number of stars\footnote{$\varpi/\sigma(\varpi)>20$ and proper motion $>3$ mas/yr from the Serpens centroid.}  within some distance range
 is divided by the number of stars within some more nearby range of distance.  For example, the dark clouds in Serpens can be seen in the ratio of the number of stars  between $500-600$ pc, $N(500-600)$, divided by the number of stars between $300-400$ pc, $N(300-400)$.  All ratios are normalized by the median ratio from regions in our map that are assumed to be free of extinction.  

\begin{figure}[!t]
\epsscale{1.35}
\includegraphics[width=0.49\textwidth, trim=60 83 40 355,clip]{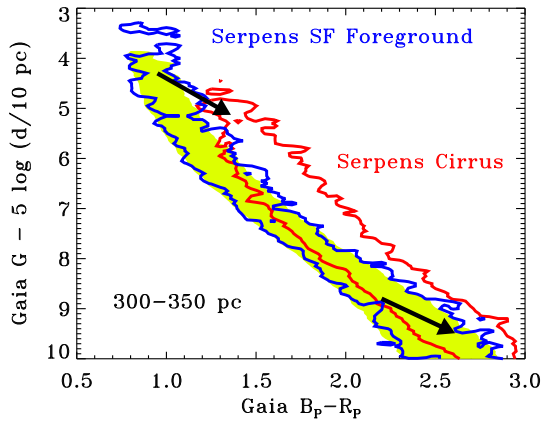}
\caption{The color-magnitude diagram of all stars from 300--350 pc with $\varpi/\delta(\varpi)>20$ and $G<19$, for Serpens Cirrus (red contour, sources within a large rectangle from $\alpha=262$ to $278$ and $\delta=-12$ to $-2$ deg.), the foreground of Serpens (blue contour), and the rest of the field.  Compared to the rest of the map, the stars in Serpens Cirrus are redder and fainter, seen in the different locations of the top of the main sequence, as expected for a population behind a cloud of dust.  Extinction vectors (arrows) are shown for $A_V=1$ (see Appendix A).}
\label{fig:hrdiag_aquila}
\end{figure}

The contours of star count ratios show the locations where extinction occurs between those two distances\footnote{The location of these dust features are robust to different choices for the expected ratio of number of stars in each distance bin, including corrections for galactic latitude.}.    The prominent  extinction features associated with the Serpens star-forming compexes are located at 400--500 pc.   
Most of the extinction to the northeast of the Serpens star-forming cluster and in some adjacent dark clouds occurs at $\sim 650$ pc, called the Aquila Rift.  Star counts at distances beyond $>1000$ pc demonstate that the Aquila Rift extends across our entire field-of-view.
    The region to the southwest of Serpens star-forming regions, here called Serpens Cirrus, has a deficit of stars between 300--400 pc, relative to the number between 30--250 pc.   The color-magnitude diagram of stars toward Serpens Cirrus   (Figure~\ref{fig:hrdiag_aquila}) confirms the $\sim 250$ pc distance to this dusty region, with a stellar locus that is offset by $\sim 1$ mag in $A_V$ from the downloaded area (see Appendix A for the calculation of extinction vectors).  The stars in front of the Serpens star-forming regions are not reddened relative to off-cloud regions, indicating the absence of any significant foreground extinction.

These comparisons provide an accurate assessment for the distance to the closest location that extinction occurs along the line-of-sight, if the extinction is sufficiently opaque.  Extinctions of $A_G<0.5$ may not be readily detectable in such star count ratios for nearby regions, but significantly alter the detected distribution of apparent magnitudes.  Such star count maps may therefore create the misperception that an optically-thin, nearby cloud is actually located at a much larger distance.

\subsection{Quantifying the distance to individual clouds}

The star count maps described above reveal regions with significant extinction.  To quantify the distance to regions of extinction, we compare the measured star counts versus distance to simulated star counts from off-cloud regions, after introducing a cloud at a given distance.  Because the astrometric accuracy depends on brightness, the populations need to be simulated to quantify the  distance and distance uncertainty to the extinction region.  Appendix B describes our simulations of star counts versus distance for a sheet of extinction at some distance.  

\begin{figure*}[!t]
\epsscale{1.3}
\hspace{-25mm}
\plotone{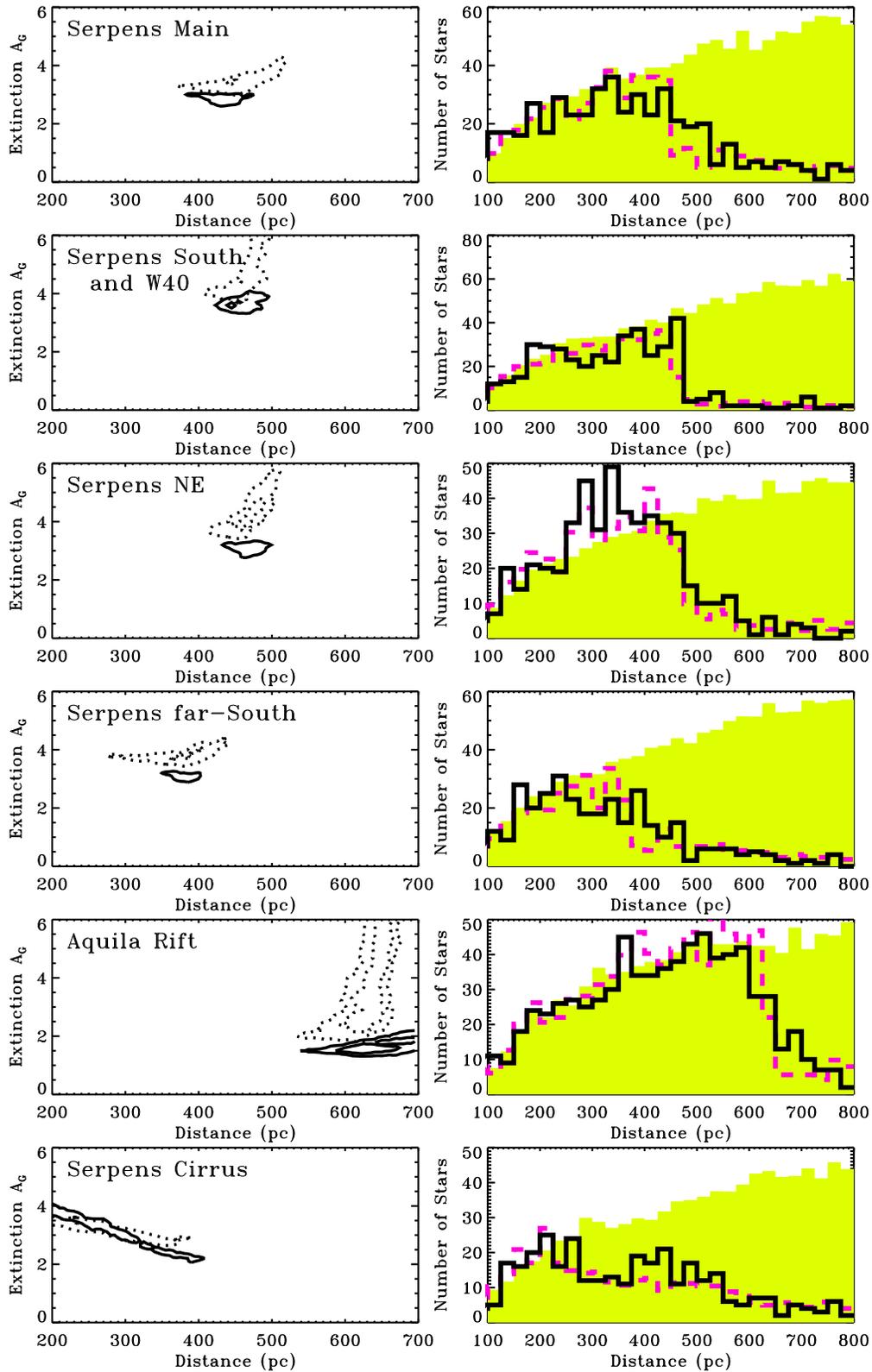}
\vspace{-40mm}
\caption{{\it Left:}  The acceptable distance and extinction parameters from the KS test between a simulated and observed region with confidence contours of 5\% and 32\%, for stars with parallaxes measured with $S/N>20$ (dotted contours) and $S/N>5$ (solid contours).  The location of the region is provided in Table~\ref{tab:dustdist}; Aquila Rift is LDN 610, while Serpens Cirrus is near LDN 462.  {\it Right:}  The star counts versus distance for the region on the left, comparing the observed (black line) and simulated (dashed purple line) populations.  The stars in this comparison all have parallaxes measured to $S/N>20$.  The simulated populated is obtained from a background population (yellow shaded region) with the same absolute galactic latitude at the target region, but south of the galactic plane.}
\label{fig:contourdist}
\end{figure*}

Table~\ref{tab:dustdist} and Figure~\ref{fig:contourdist} present our distance measurements to dust extinction for 12 circular areas on the sky.  The dark clouds associated with young star clusters all have distances that are consistent with the measured distances to the young clusters (see \S 3).  
Along the Aquila Rift, the optically-thick clouds are located at $\sim 500$ pc to the southwest and $\sim 600$ to the northeast of Serpens.  
Serpens Cirrus is located closer to us than the Serpens star-forming clouds, as suggested in \S 4.1.  
The star counts to Serpens South and Serpens Main appear to have a deficit at $\sim 250$ pc,
which is well explained by simulating Gaia data from off-cloud populations for a single, 
optically-thick cloud at $\sim 450$ pc and should not be interpreted as indicating the 
presence of foreground dust at the distance of Serpens Cirrus (see also 
Figure~\ref{fig:hrdiag_aquila}).  Stars that are in the foreground of the cloud may have a 
measured parallax (after accounting for uncertainties in parallax) that leads to a distance 
behind the cloud.  Most stars that are physically behind the cloud are too faint to be 
detected by {\it Gaia}, and therefore no corresponding population of background stars has 
measured parallaxes that would place them in front of the cloud.

These distances are calculated by estimating the probability that the star counts for each region are from the same sample as the extinction-corrected off-cloud population.  The probability is evaluated with a Kolmogorov-Smirnov test, for simulations with distance $d$ and extinction $A_G$ in a grid with a spacing of 5 pc and 0.1 mag.  The star counts used for the Kolmogorov-Smirnov test have parallaxes with distances from 100--800 pc, measured with $\varpi/\sigma(\varpi)>20$.  The adopted  distance in Table~\ref{tab:dustdist} corresponds to the grid point with the highest probability in the Kolmogorov-Smirnov test.
The contours in Figure~\ref{fig:contourdist} show the parameter space that can be ruled out at $68\%$ and $95\%$ confidence.  The $d_{\rm min}$ and $d_{\rm max}$ are then evaluated as the minimum and maximum distances that cannot be ruled out at the 95\% confidence level.  These 12 regions are selected in part because their best-fit extinction is $A_G>2$ mag, which generates tighter constraints on the distance than if the extinction were lower; a few areas within our downloaded dataset have high extinction but were also excluded.

The parent population for these measurements is obtained from regions that are thought to be unaffected by extinction and are located at a similar galactic latitude (within $0.5^\circ$ of the centroid), but south of the galactic plane.  Most extinction in the downloaded Serpens-Aquila area occurs at north galactic latitude.  In simulations with off-cloud populations located in north of the galactic plane or separately in annuli around each region, the distances are similar but extinctions are smaller.

The distance to a dark cloud corresponds to the front of that cloud, where the stellar density initially drops.  Most dust structures in our image extend $\sim 1$ deg ($\sim 7$ pc)in diameter on the sky and are therefore not deep along our line-of-sight.  The assumption that these dust clouds are located at a single distance is reasonable, unless the structures are 1-dimensional filaments aligned along our line-of-sight.  However, each region may include multiple clouds located at different distances, and the clouds themselves are likely filamentary.

\begin{table*}
\begin{center}
\caption{Distances to extinction regions$^a$}
\label{tab:dustdist}
\begin{tabular}{clcccccc}
Number & Name & RA & Dec &  $d$ (pc) & $d_{\rm min}$ (pc) & $d_{\rm max}$
(pc) & $A_G^c$ \\
\hline
1 & Serpens Cirrus &    271.00 &    -4.70 &   235 &   200 &   405 &    3.4\\
2 & LDN 486$^d$&    272.00 &    -3.30 &   515 &   430 &   575 &    2.7\\
3 & LDN 500$^d$ &    273.50 &    -3.30 &   495 &   200 &   585 &    2.6\\
4 & LDN 505$^d$ &    275.50 &    -3.00 &   445 &   365 &   480 &    3.3\\
5 & Serpens far-South &    277.20 &    -3.50 &   370 &   350 &   400 &    3.7\\
6 & Serpens Main       &    277.30 &     0.60 &   445 &   385 &   475 &    3.4\\
7 & Serpens South/W40 &    277.60 &    -2.05 &   460 &   425 &   495 &    4.4\\
8 & Serpens               &    278.50 &    -0.80 &   415 &   375 &   470 &    3.0\\
9 & Serpens NE  &    279.50 &     0.20 &   465 &   435 &   495 &    4.0\\
10 & LDN 603$^d$        &    281.50 &     1.50 &   585 &   480 &   590 &    2.0\\
11 & LDN 566 &    282.30 &    -2.60 &   590 &   540 &   695 &    2.0\\
12 & LDN 610    &    282.40 &     2.80 &   630 &   540 &   695 &    2.7\\
 \hline
\multicolumn{8}{l}{$^a$Calculated for projected circles with radii of $0.7$ deg from the listed center}\\
\multicolumn{8}{l}{$^b$$d_{\rm min}$ and $d_{\rm max}$ are minumum and maximum distances}\\
\multicolumn{8}{l}{~~!~that cannot be ruled out at 95\% confidence, see Figure~\ref{fig:contourdist}.}\\
\multicolumn{8}{l}{$^c$Extinction values are an average over the full region and are not reliable.}\\
\multicolumn{8}{l}{$^c$LDN association is approximate.}\\
\end{tabular}
\end{center}
\end{table*}

\section{Discussion and Speculation}

The {\it Gaia} DR2 view of the Serpens star-forming regions and the Aquila Rift reveals a complex set of star clusters and molecular clouds.  The young stars are distributed across five different sub-clusters, three of which are detected by {\it Gaia} DR2 and have measured distances of 383--478 pc.  The subclusters that are furthest apart, Serpens far-south and Serpens Northeast, are separated by 35 pc in projected space but by 95 pc along our line of sight.  These clusters are likely part of the same cloud complex, since the molecular gas is located at similar velocities \citep{dame85,dame01,davis99}.  The Serpens star-forming cloud complex is therefore elongated along our line of sight, and may be a branch of the much larger set of molecular clouds in the Aquila Rift, which has a distance of 
600--700 pc to the northeast of the Serpens star-forming regions and$\sim 500$ pc to the southwest of the active star-formation.

Some light extinction also occurs to the southwest of the main Serpens complexes at $\sim 250$ pc, called Serpens Cirrus.  The distance to Serpens Cirrus confirms the accuracy of distance measurements to dust extinction from the maps of \citet{straizys03}. However, this distance should not be applied to any known clusters of young stars, consistent with the conclusions of \citet{prato08}.  Serpens Cirrus is not entirely devoid of ongoing star formation; the dark cloud L483 
hosts a protostar \citep[e.g.][]{tafalla00} and with an approximate distance of 250 pc seems to be associated with Serpens Cirrus. Some candidates identified by \citet{dunham15} in the southwestern-most region of their map, around ($271$,$-4.5$), may also be young stars.
The separation in distance between the main Serpens star-forming clouds and the Serpens Cirrus is also  consistent with the 3D extinction maps of \citet{lallement14}.

The distances listed here are calculated from parallax measurements that have systematic uncertainties of $\sim 0.1$ mas \citep{luri18}.  The accuracy of VLBI measurements therefore exceeds that of {\it Gaia} DR2.  
The distance to two sources in Serpens Main\footnote{The VLBI distance to Serpens Main excludes the star GFM 65, which has a distance and proper motion that are highly discrepant from the Serpens Main cluster, indicating either that the observational uncertainties are underestimated for this source, that it is a remarkable runaway, or that it is a non-member.  The distance would place the star in front of the cloud.} of $436\pm7$ pc, as obtained from VLBI measurements of maser emission \citep{ortiz17ser}, is consistent with our distance of  $438 \pm 11$ (statistical $1\sigma$ error) $\pm 19$ pc (systematic) to cluster members and $445^{+30}_{-60}$ pc (error bar corresponds to 95\% confidence interval) to the cloud extinction.  The VLBI distance to four sources in W40 and Serpens South of $436\pm7$ pc is also consistent with the distance of $460\pm35$ pc to the dust extinction to W40 and Serpens South.  The number and color-magnitude diagram of stars in this region at $\sim 300-400$ pc indicates a lack of optically-thick dust within those distances, thereby excluding the possibility that Serpens South is located in the foreground. 

Only five stars have astrometry from both {\it Gaia} and VLBI, and three of those stars have high astrometric noise, so we cannot evaluate systematic differences.  The average proper motion of Serpens Main calculated here is consistent with the VLBI proper motions of GFM 11 and EC 95.  Of the four VLBI astrometric measurements in in W40 and Serpens South, three have proper motions that are offset by a few mas/yr from the motion of the other Serpens clusters, consistent with a possible offset of $\sim 5$ \kms\ in  proper motion of faint candidate members (see \S 5.3.2).

The distances to the Serpens star-forming clouds are consistent with the distances estimated by \citealt{green15} from 3D dust maps obtained with Pan-STARRS \citep[see also][]{green18}.  However, the {\it stilism} \citep{capitanio17} map of extinction suggests that the Serpens clouds are located at $\sim 700$ pc, perhaps because of cloud confusion.  Our distances are also remarkably consistent with the supposedly-crude $\sim 440$ pc photometric distance to HD 170634, located in Serpens Main, by \citet{racine1968} and \citet{strom1974}.

\subsection{Evaluating Selected Past Membership Surveys}

{\it Gaia} astrometry allows us to evaluate the optically-bright candidate members of Serpens that have been identified by past surveys.  
Table~\ref{tab:members} lists the number of objects in selected papers with a {\it Gaia} match\footnote{Including all {\it Gaia} DR2 targets with astrometry, with no limit on $G_P$ but restricted to targets with excess astrometric noise $<2$ mas}, with statistics following the classification of each match as either a robust non-member or as consistent with kinematic membership. The contamination rates described here are generally measured to stars that are not located deep within the clouds.  In regions with high densities of members, the contamination rates are likely much lower.  On the other hand, in regions with low densities of members but with dark clouds, the contamination rates are likely underestimated because background objects will be too faint for detection by {\it Gaia}.

Non-membership is assessed if the proper motion is more than 2 mas/yr ($4\sigma$) away from the Serpens centroid or if the distance is $2\sigma$ away from the 370--490 pc range.  Stars that have kinematics consistent with membership have proper motions within those ranges.  Consistency with the Serpens distance is ignored if the parallax uncertainty is larger than $\sigma(\varpi)=0.3$ mas, although inconsistency may be assessed for any parallax uncertainty.  
{\it Gaia} stars are considered a match if they are within $3^{\prime\prime}$ of a {\it Spitzer} mid-IR position and $1^{\prime\prime}$ of a near-IR or X-ray position.  Most stars that are listed in catalogs but not matched with {\it Gaia} targets are optically-faint, likely because they are located within or behind molecular clouds, because they are embedded in a dense envelope, or both.

\begin{figure*}[!t]
\includegraphics[width=0.49\textwidth, trim=0 0 0 0,clip]{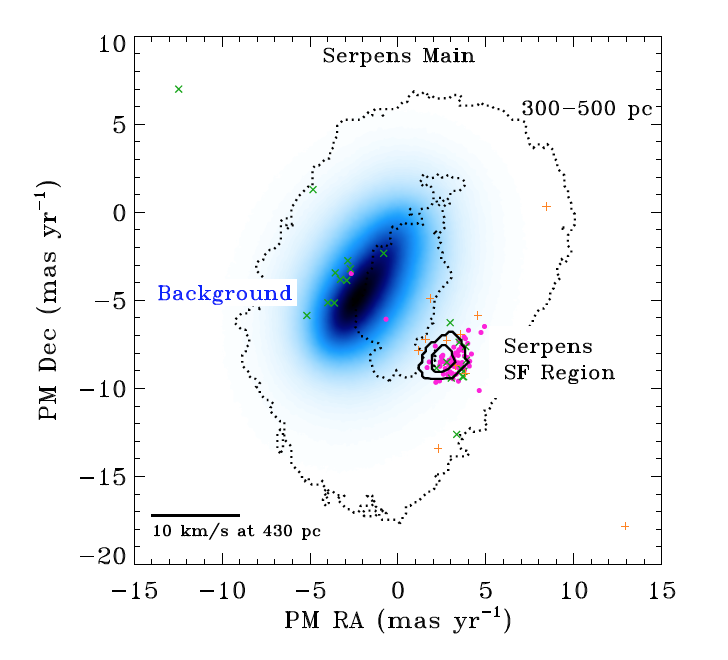}
\includegraphics[width=0.49\textwidth, trim=0 0 0 0,clip]{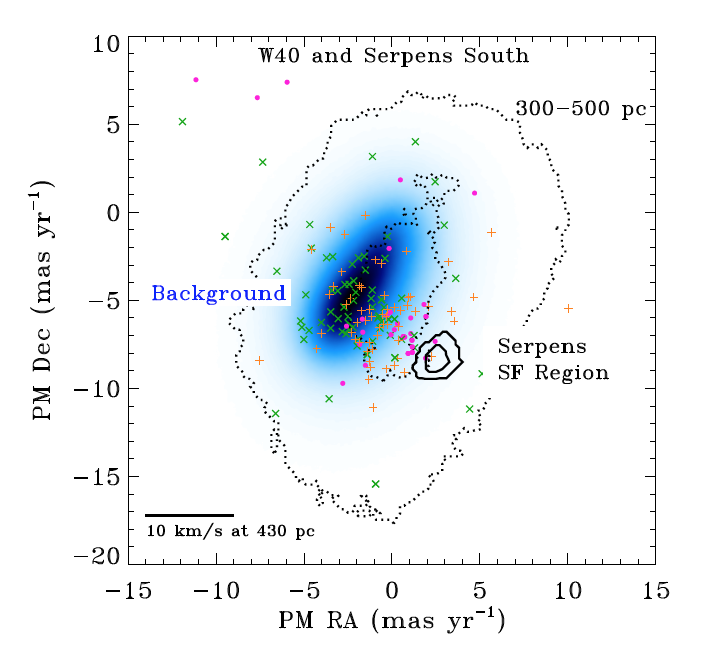}
\includegraphics[width=0.49\textwidth, trim=0 0 0 0,clip]{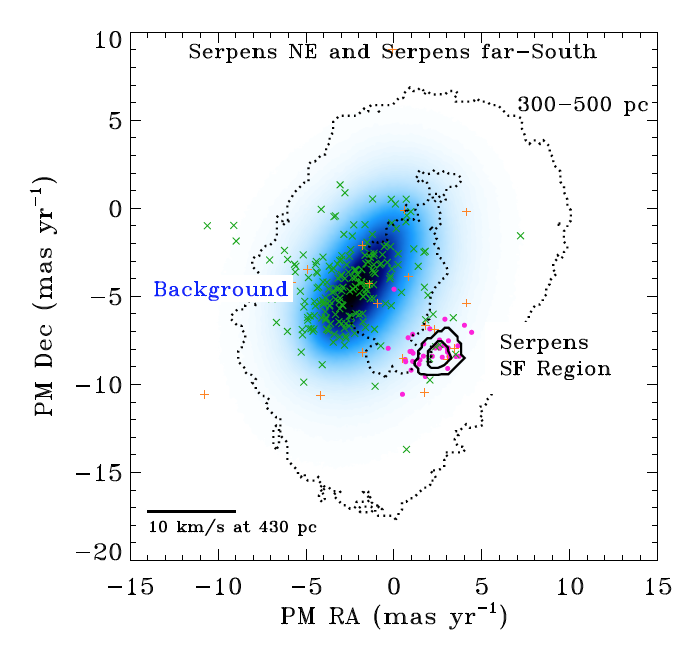}
\hspace{2mm}
\includegraphics[width=0.49\textwidth, trim=0 0 0 0,clip]{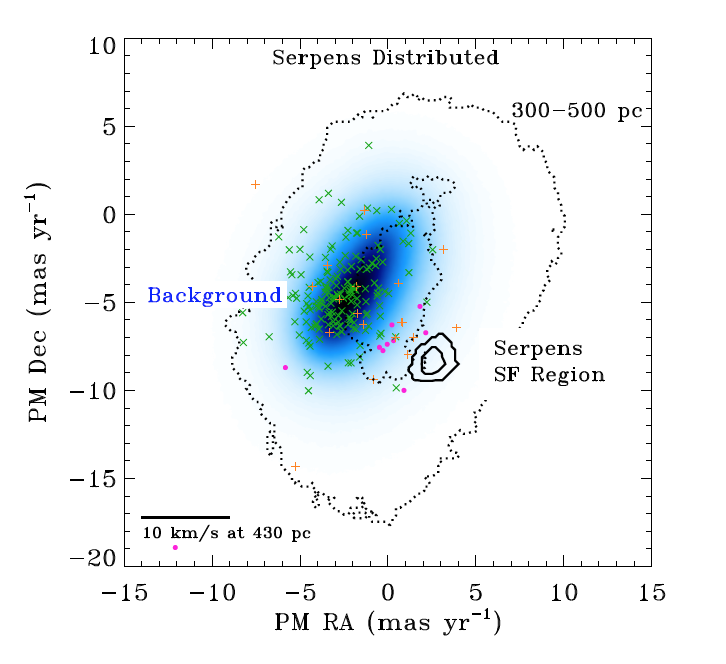}
\caption{The proper motions of candidate young stars identified by \citet{dunham15} in the Serpens Main (top left), W40 and Serpens South (top right), Serpens Northeast and Serpens far-South (combined into one plot, bottom left), and elsewhere in Serpens outside of these regions (bottom right), compared with the proper motions of the young stars in Serpens star-forming regions (solid contours at 0.2 and 0.5 times the peak value), all stars in our region with distances of 300--500 pc (dotted contours), and all background objects with distances $>1000$ pc (shaded blue contours), and objects with parallax errors that lead to uncertain membership (orange pluses).  The candidate young stars that have {\it Gaia} DR2 distances  consistent with the Serpens star-forming complex (purple circles) are clustered around the Serpens star-forming region, confirming membership, while objects with other distances (green x's) have proper motions that correspond to background stars.  Many stars in Serpens South and W40  are clustered in proper motion a few mas/yr away from the main Serpens SF regions, which may be interpreted either as consistent with the field population at 300--500 pc or as sub-cluster with a large scatter in proper motion.}
\label{fig:pmdunham}
\end{figure*}

The evaluation of memberships described here is based primarily on the census of Serpens from {\it Spitzer} maps of \citet{dunham15}, which cover a wide area on the sky\footnote{\citet{dunham15} splits the population into Serpens, which includes Serpens Main and Serpens Northeast, and Aquila, which includes Serpens South, Serpens far-South, and regions along the Aquila Rift to the southeast of Serpens.  We divide these regions visually into additional sub-regions, see Table~\ref{tab:members}.}.  Since the \citet{dunham15} catalog establishes membership for most of the Gould Belt, this evaluation helps to inform us of the reliability of different aspects of that catalog.  Serpens Main and Serpens South have additional membership catalogs that are also evaluated here.  For regions with optically-thick dust, the {\it Gaia} matches are optically bright and are therefore more likely to be foreground non-members than either members or background contaminants, so contamination rates for the youngest regions are unreliable.
{\it Gaia} kinematics are also unable to probe the protostellar population, though the contamination rate of that population is likely to be low; \citet{heiderman15} confirmed that 90\% of protostar candidates in the \citet{dunham15} catalog have envelopes.

Most previously known stars with {\it Gaia} kinematics consistent with membership are located in Serpens Main.  The {\it Gaia}-matched Class II objects from 
\citet{dunham15} have a membership probability of 86\%, while the evolved
(Class III) objects have a 50\% probability of membership.  Other IR and X-ray surveys have similar contamination rates.  The 47\% contamination rate for 74 {\it Gaia}-matched objects in the \citet{eiroa08} catalog is higher than expected.  Inspection of these contaminants indicate that many were classified as candidates from near-IR photometry and lack spectroscopic confirmation.

\begin{figure}[!t]
\epsscale{1.0}
\vspace{5mm}
\hspace{-3mm}
\plotone{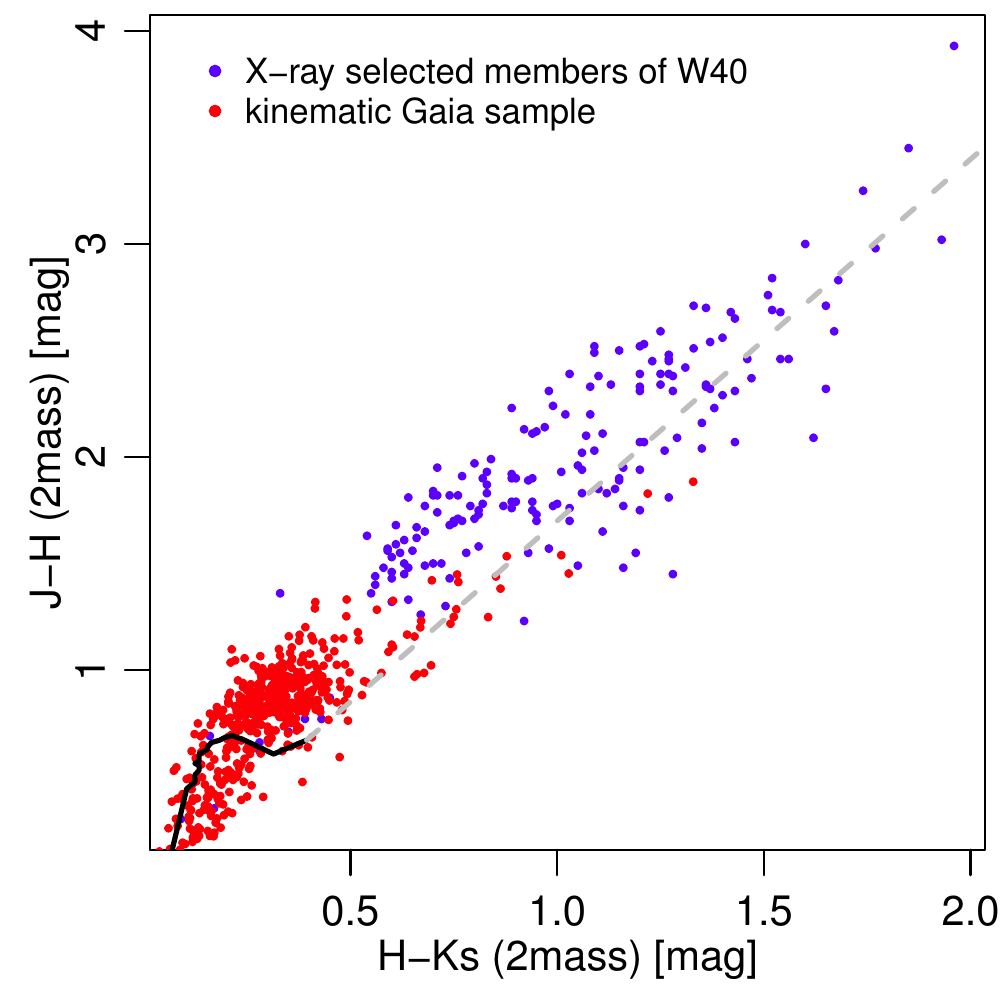}
\caption{2MASS color-color diagram for the kinematically selected sample (red) and X-ray selected cluster members from W40 \citep[blue, from][]{kuhn13}, with a 1 Myr isochrone from \citet{siess00}.  A higher fraction of W40 members than {\it Gaia} kinematically-selected members have a K-band excess from an inner disk.  The {\it Gaia}-selected sample is only the surface population of slightly older, optically-bright stars within a large complex that harbors a rich population of protostars in several sites of ongoing star formation.}
\label{fig:ccd}
\end{figure}

For Serpens South and W40, very few sources that have been previously identified as candidate members can be evaluated with {\it Gaia}. The rejection rates from samples of disks and X-ray emitters \citep{kuhn10,dunham15,getman17,winston18} with {\it Gaia} matches are 60\%, while the rejection rate for \citet{povich13} is 80\%. However, many of the objects in this region are clustered in proper motion and are offset by 5 \kms\ from the proper motion centroid of the three main optical clusters.  These objects may be young, if the proper motion of Serpens South and W40 are offset from the rest of the Serpens complex (see further discussion in \S 5.3.2).

In most locations beyond the dense clusters, the {\it Gaia} matches to the \citet{dunham15} catalog demonstrate that the candidates are non-members, most with kinematics consistent with the background population.  The classification of stars as background AGB stars by \citet{dunham15} are almost all confirmed.  Most evolved stars outside of Serpens Main and Serpens South are also classified here as background objects; this category was intended to include diskless (class III) young stars with an expected contamination rate of 25--90\%.  The protostars in Serpens Northeast and Serpens far-South are likely robust.  The one protostar and other matches to the southwest of the main Serpens clusters may be coincidental.  

These contamination rates in the Spitzer-based population of 
Serpens and the Aquila Rift are likely not significant for 
some other regions evaluated by \citet{dunham15}. 
 The background population of AGB stars is much higher for Serpens, 
 seen against the galactic plane,
than for other nearby regions in the Gould Belt, which have higher galactic latitude. 
However, the Spitzer-based census of the Lupus V and VI star-forming region suffers from similar contamination of background stars, when evaluated with {\it Gaia} DR2 astrometry \citep{manara18}.   The contamination in these populations will affect ratios of pre-main sequence stars in different stages of evolution.

\subsection{$K$-band Excess Disk Fractions of the Gaia members}

For young pre-main-sequence stellar populations, disk fractions may be used as a proxy for relative stellar ages \citep[e.g.][]{haisch01,hernandez08,2009AIPC.1158....3M,fang13,2018MNRAS.477.5191R}.  In this analysis, we use 2MASS JHK$_s$ photometry \citep{cutri03} to directly compare the inner disk fraction of the {\it Gaia}-selected kinematic members of Serpens to the disk fraction analysis of W40 by \citet{kuhn10}.  Excess emission in the $K_s$ band is used here as a proxy for emission from hot inner disks \citep[see, e.g.,][]{meyer97}.  Although less sensitive  than a mid-IR survey, 2MASS photometry covers the full region (unlike Spitzer), has fewer problems with blending and saturation in W40, and may be directly compared to a previous analysis of W40 using the same methodology.

Figure~\ref{fig:ccd} shows kinematically-selected sources from this work and X-ray-selected sources from \citet{kuhn10} on a $H-K_s$ versus $J-H$ color-color diagram.  Following \citet{kuhn10}, the 1~Myr \citet{siess00} isochrone is drawn and a reddending vector from 0.2~$M_\odot$ is used as an approximate separation between sources with and without $K_s$-band excess. The majority of the stars within W40 are more deeply embedded than the population seen by {\it Gaia}.  In W40, the $K_s$-excess fraction of $\sim$50\% (after correcting for selection effects), while for the Gaia-selected sample the K$_s$ excess fraction is $<4$\%. The {\it Gaia} sources that do lie to the right also tend to be subject to higher extinction than the majority of the {\it Gaia} sources.

The lower inner disk fraction for the {\it Gaia} kinematic sample is consistent with the expectation that the optically-bright populations are older stars, while the younger stars and protostars are located deep within the molecular cloud.  This radial age gradient is often seen in star-forming regions \citep{getman14,getman18,jose16}, either because  older stars drift outward or because the molecular material rapidly disperses after forming stars (see \S 5.3.3).
This comparison of near-IR disk fractions are also influenced by selection effects, since local extinction from circumstellar disks may hamper the sensitivity of {\it Gaia}, and by differences in applying this technique to samples with very different extinctions and mass distributions (see also \citealt{soderblom14} for a description of the uncertainties in using disk fractions for age estimates).   An evaluation of disk presence with mid-IR photometry would yield higher disk fractions. Nevertheless, the 4\% fraction of stars with inner disks is much lower than would be expected for a very young region, such as W40, and may indicate that the optically bright population has an average age of 5--10 Myr.  The more embedded regions that were not detected by {\it Gaia} have much younger ages. 

\begin{figure*}[!t]
\includegraphics[width=0.49\textwidth, trim=45 60 45 320,clip]{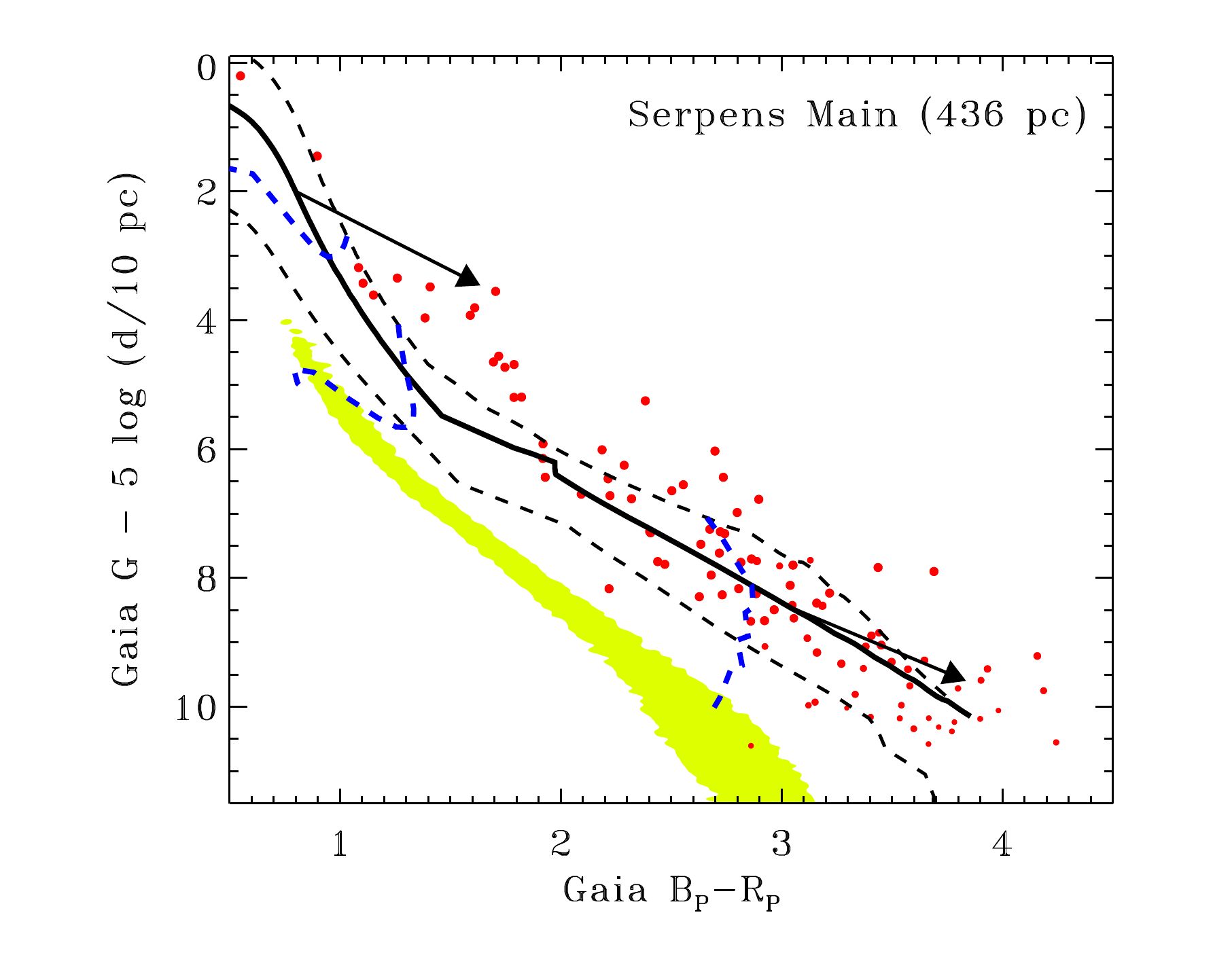}
\includegraphics[width=0.49\textwidth, trim=45 60 45 320,clip]{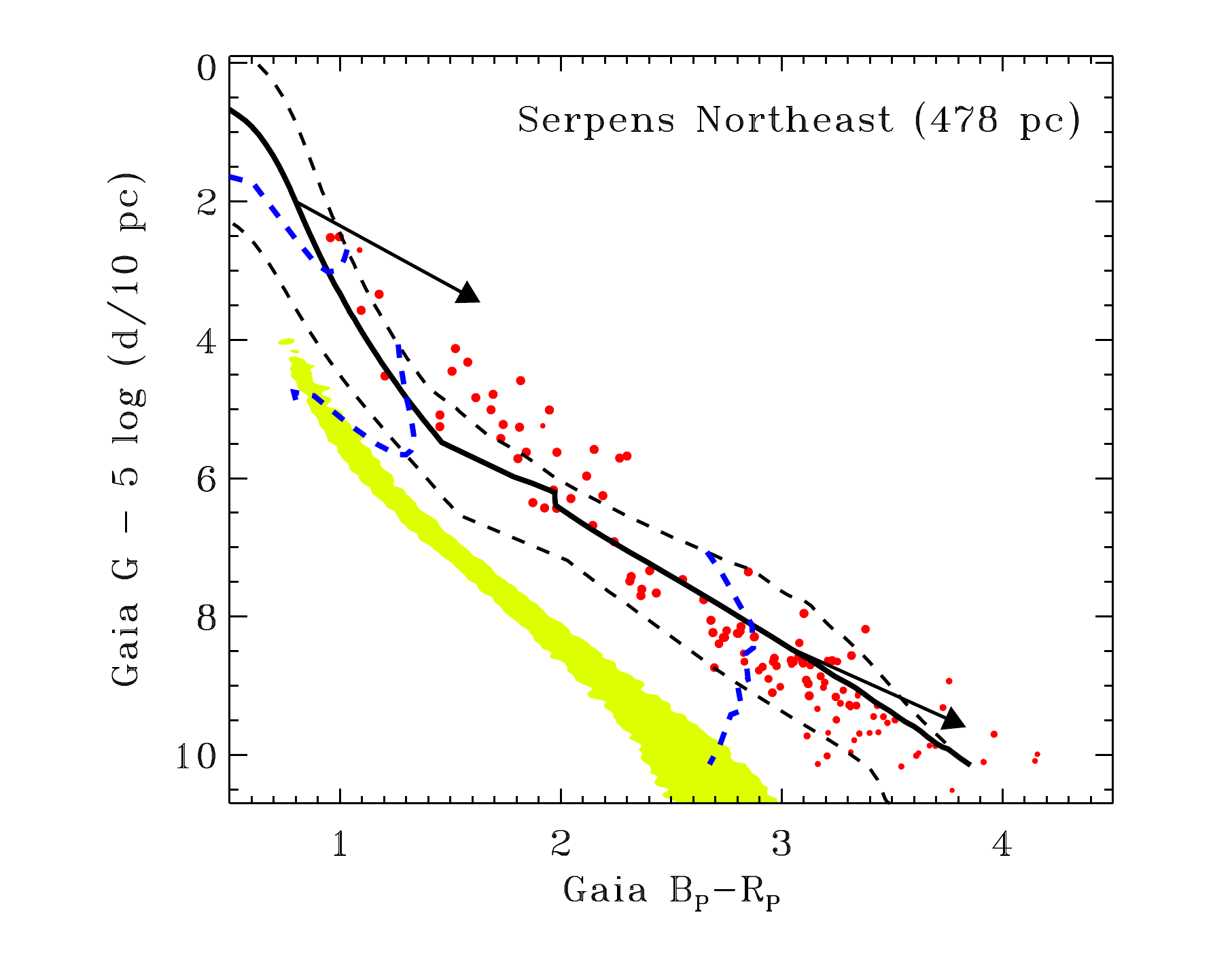}
\includegraphics[width=0.49\textwidth, trim=45 60 45 320,clip]{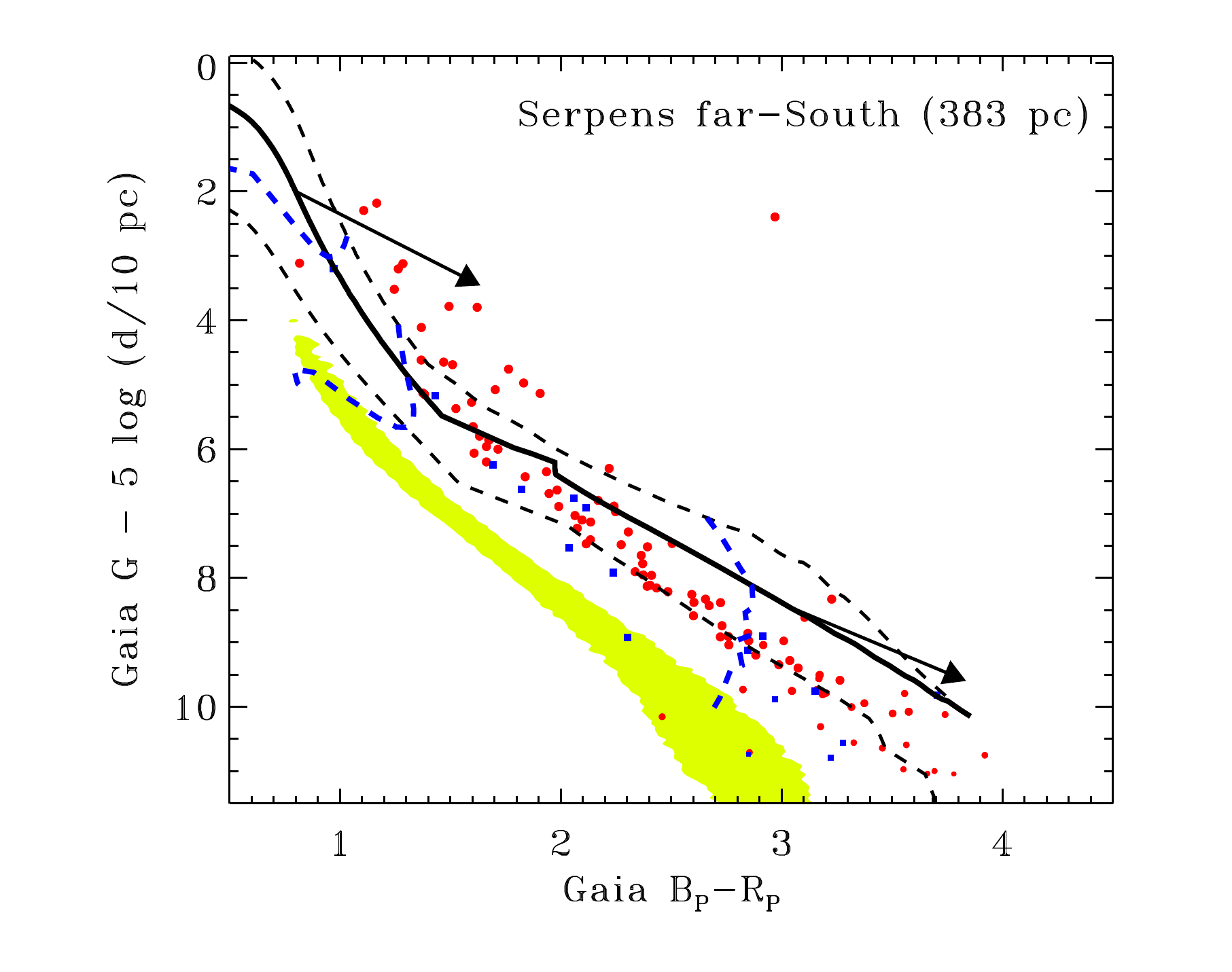}
\hspace{2mm}
\includegraphics[width=0.49\textwidth, trim=45 60 45 320,clip]{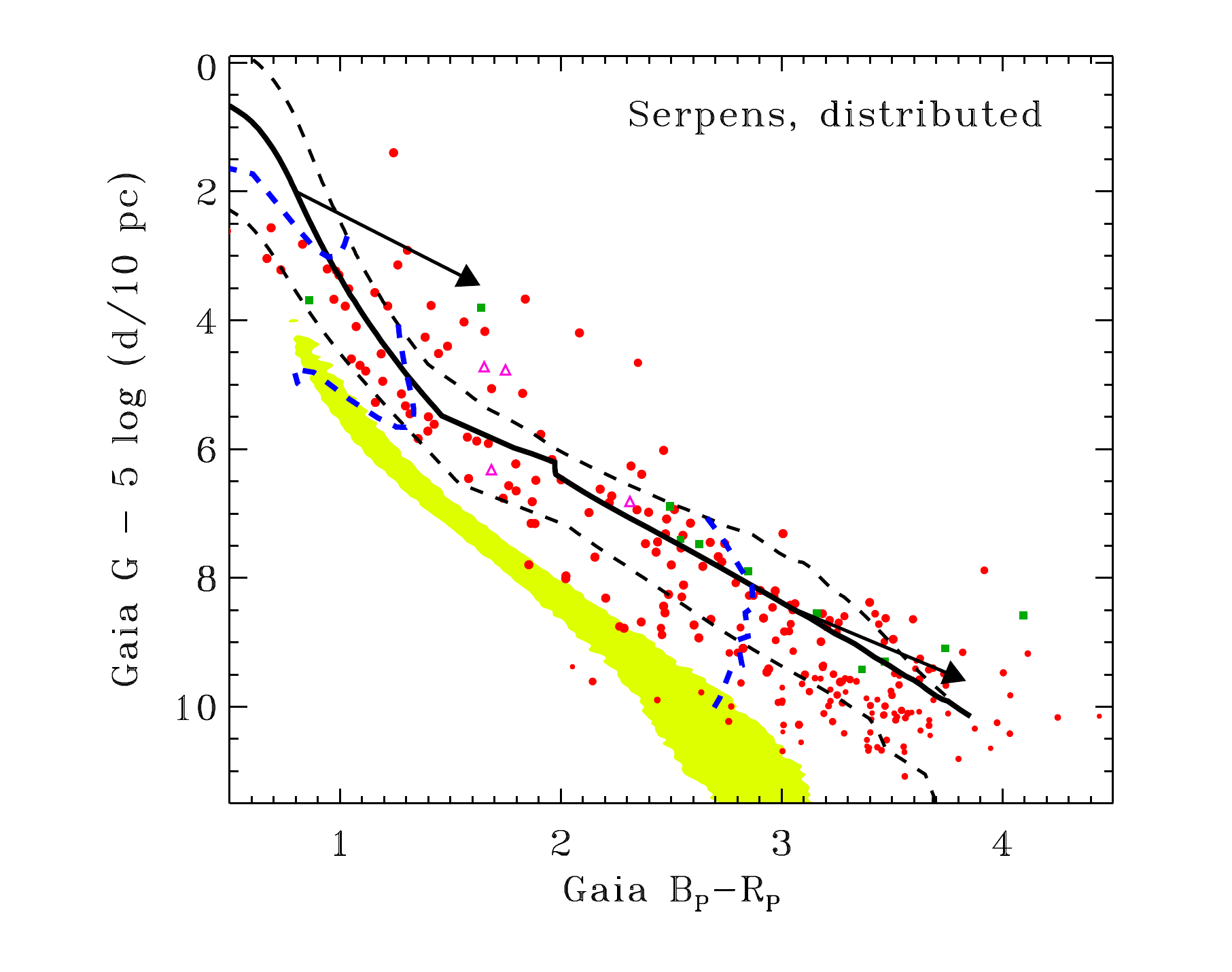}
\vspace{-5mm}
\caption{Color-magnitude diagrams of young clusters in Serpens, compared with unreddened isochrones of 1, 2.5, and 10 Myr and evolutionary tracks for stars with 2.0, 1.0, and 0.3 M$_\odot$ \citep{bressan12,marigo17}.  The size of points indicate the accuracy of the parallax measurement, from $\varpi/\sigma(\varpi)>20$ (largest) to $\varpi/\sigma(\varpi)=5$ (smallest).  The distance-corrected magnitudes are calculated assuming a single cluster distance, as listed.  The lower left panel includes the small group at $\alpha=276$, $\delta=-6$ (blue squares).  The lower right panel includes all high-probability members that are not associated with any of the three main optical clusters, including the LDN 673 group (407 pc, green asterisks) and the FG Aql group (408 pc, green squares); individual stellar distances are used for the distributed population.  The main sequence (yellow shaded region) is calculated from all stars in our field between 50--250 pc and with $\varpi/\sigma(\varpi)>20$.}
\label{fig:hrdiags2}
\end{figure*}

\subsection{Describing the clusters}

The young stars in Serpens are preferentially located in a few clusters.
Figure~\ref{fig:hrdiags2} shows color-magnitude diagrams for the three optical 
clusters and for the distributed population of young stars in the region.  The
magnitudes are corrected for the average distance to each cloud but are not corrected for extinction.  Individual stellar distances are used for the distributed population.

Almost all candidate members are located above the main sequence and are consistent with the pre-main sequence locus, despite a selection that did not
include photometry.  These candidates therefore have a very high membership
probability, even without spectroscopic confirmation.  Table~\ref{tab:starprops}
provides the {\it Gaia} DR2 astrometry and photometry for these high-probability objects.  A complete (rather than a statistical) sample of members
would require  spectroscopic confirmation of stars with low-to-modest probability of membership.

The stellar positions of likely members are compared to the unreddened PARSEC isochrones \citep{bressan12,marigo17} using the BT-Settl colors \citep{allard14}.  
If we infer a typical extinction of $A_V\sim2$ mag (neglecting differential reddening within a cluster and the effects of accretion), then this {\it Gaia} DR2 identification of members is sensitive to $\sim 0.2-0.3$ M$_\odot$, with a completeness that is high in low-extinction regions and for higher masses.  The estimated cluster ages for this extinction range from $<1$ Myr for Serpens Main to $\sim 3$ Myr for Serpens far-South and the distributed Serpens population.  These ages are younger than expected for the ($K$-band excess) disk fraction of 4\%.  Resolving this tension will require analyses of individual stars, which is beyond the scope of this paper, and the resolution of long-standing challenges with ages of young stars \citep[see review by][]{soderblom14}.  However, even if the oldest populations have ages of 10 Myr, the challenges described below are still robust, just scaled to an older age.

\begin{figure*}[!t]
\epsscale{1.1}
\plotone{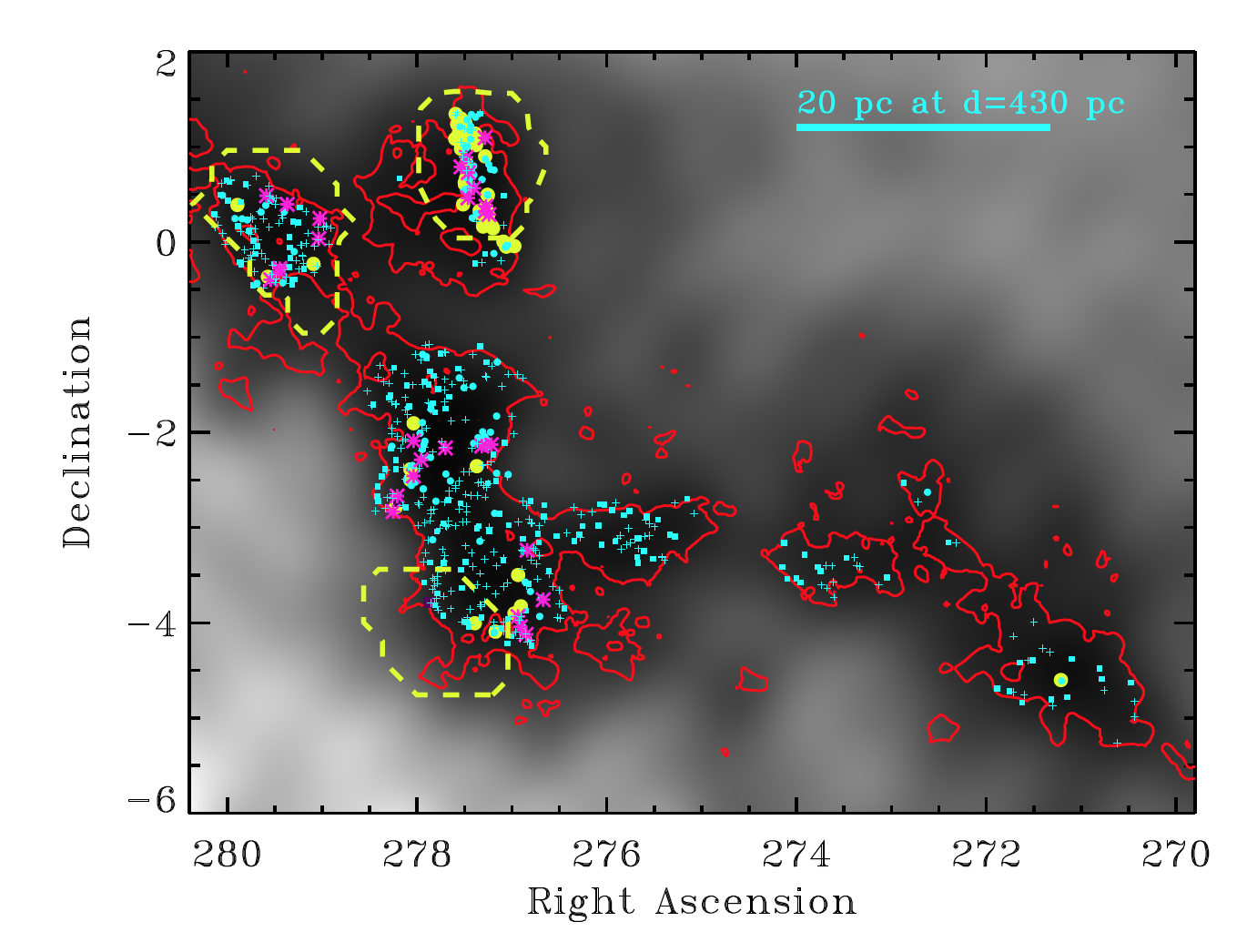}
\caption{Objects from the \citet{dunham15} {\it Spitzer} catalog of candidate members of the Serpens-Aquila star forming complexes, plotted against the starcounts map from Figure~\ref{fig:largemap}.  The yellow circles (disks and protostars, as classified by \citealt{dunham15}), purple asterisks (diskless members), and yellow squares (AGB stars) have distances and proper motions consistent with membership, with proper motions within 4 mas/yr of the Serpens centroid and distances within $2\sigma$ of 380--480 pc.  Most of the evolved (diskless, shown as small pluses) candidates and the candidates identified as background AGB stars (small squares) by \citet{dunham15} are background AGB stars. 
The blue objects (with classifications the same shape as above, though smaller) are likely non-members.  The objects identified as disks and protostars (red circles) are almost all located in one of the major Serpens clouds.   The confirmed members are tightly confined to four specific regions of star formation; virtually no candidates from \citet{dunham15} are identified as members in the extended regions of what was then termed the Aquila Rift.  The {\it Spitzer} observations focused on regions with $A_V>5$ (red contours), as measured by \citep{cambresy99}.  The three optical clusters identified in this paper are shown in yellow contours.}
\label{fig:mapdunham}
\end{figure*}

\subsubsection{Serpens Main}
Serpens Main was the first young sub-cluster in this region to be systematically evaluated \citep{eiroa92}.  The dust extinction distance of $445^{+30}_{-60}$ pc is consistent with the stellar distance of 438 pc and the VLBI distance of 436 pc from \citet{ortiz17ser}.  The HR diagram reveals a luminous population, indicating a young ($\sim 1$ Myr) age, and a large dispersion in luminosity expected for such a young cluster \citep[see discussion in][]{soderblom14}.

The cluster is rich in both protostars and disks, with the protostars concentrated in small regions within this cloud.  If the {\it Gaia}-based contamination rate
applies to the disks and diskless populations from \citet{dunham15}, the protostar/disk fraction would be $\sim 0.42$ and the disk/diskless fraction would be $\sim 5.8$.  This sub-cluster is evolved enough for many members to be optically-bright.  No gradient in proper motion is detected, so most of the visible stars have not traveled far from their birth location.

\begin{figure}[!t]
\epsscale{1.1}
\includegraphics[width=0.49\textwidth, trim=60 70 40 295,clip]{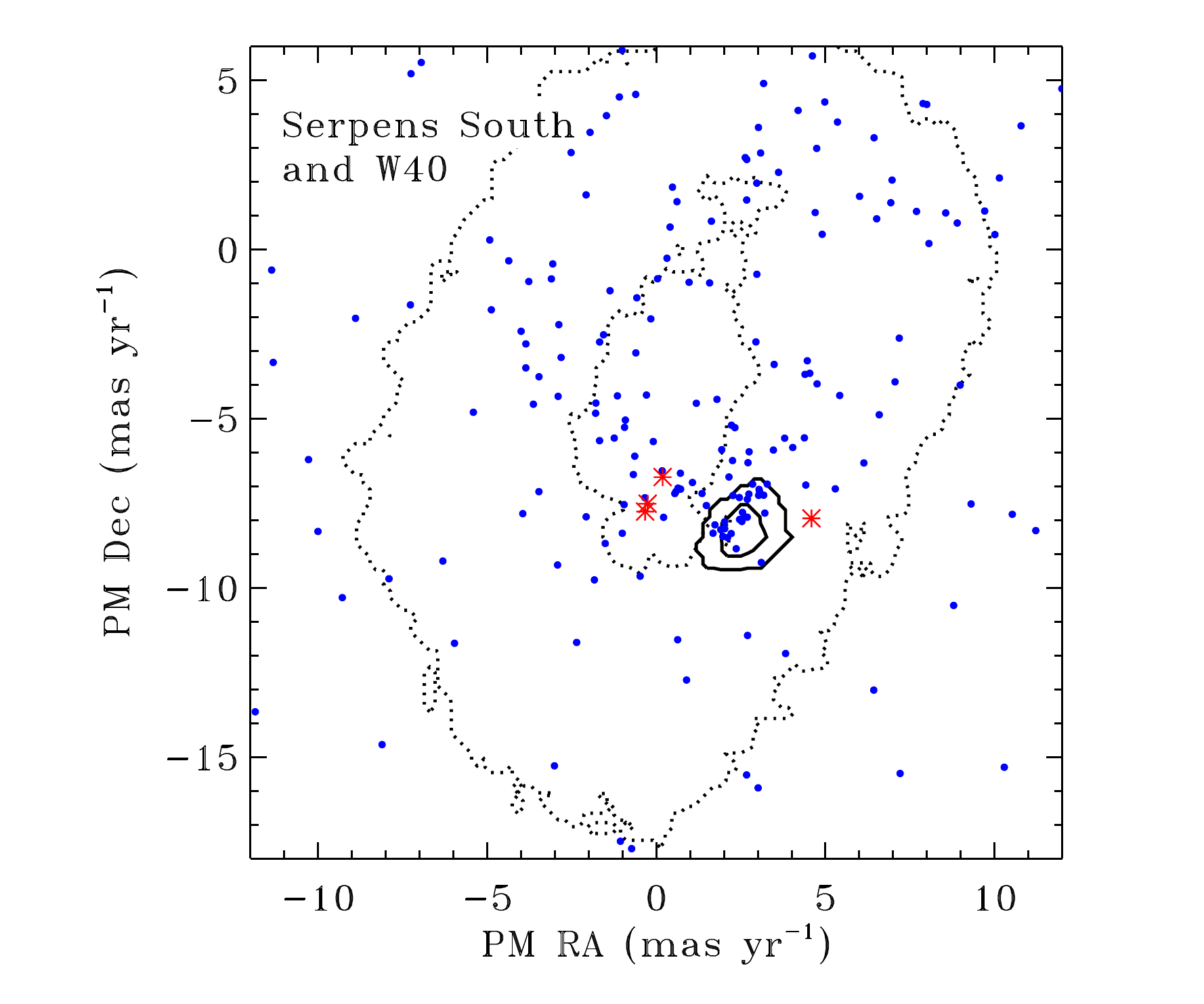}
\caption{Proper motions for stars brighter than $G<19$ located visually within Serpens South and with parallaxes indicating distances of 390--490 pc (blue circles), compared with four proper motions of stars in W40 and Serpens South as measured by VLBI (red asterisks).}
\label{fig:pm_serpsouth}
\end{figure}

\subsubsection{Serpens South and W40}

W40 and Serpens South is the most populous region in the Serpens region.  However, the protostars in Serpens South are located deep within the molecular cloud, which is why it was not recognized as such an active region of star formation until mid-IR surveys \citep{gutermuth08ser}.  \citet{dunham15} identify 374 IR-excess sources (protostars and disks), while  \citet{povich13} and \citet{mallick13} identified 983 and 1202 likely members from 2MASS and IRAC analyses, respectively\footnote{The \citet{mallick13} sample is a superset of the \citet{povich13} sample.  \citet{dunham15} required a detection in the MIPS 24 $\mu$m band, while \citet{povich13} and \citet{mallick13} required {\it Spitzer} detections only in IRAC 3.6 and 4.5 $\mu$m images.   Many additional protostar candidates have been identified with {\it Herschel} \citep{bontemps10,konyves15}.  All samples produce targets that are clustered along the filaments associated with Serpens South and W40.}.  The total population of Serpens South and W40 may be similar to the total membership of other Serpens clusters, depending on the contamination rate.

The combined W40 and Serpens South region has a VLBI parallax distance of 436 pc \citep{ortiz17ser}.  The optically-thick extinction to this combined region occurs at $\sim 460\pm 35$ pc.  We therefore conclude that the W40 and Serpens South clusters are likely located at the same distance, as inferred by \citet{ortiz17ser}, so that their VLBI parallax distance should be adopted for both Serpens South and W40 \citep[see also an independent analysis of {\it Gaia} DR2 astrometry by][]{ortiz18ser}.  Only a few candidate young stars in this area are detected by {\it Gaia} DR2, with typical distances of $\sim 410$ pc.  These stars may be biased to a distributed population of young stars in the foreground of the dark cloud.

Objects in \citet{povich13} and \citet{dunham15} with spatial locations and parallaxes consistent with Serpens membership have proper motions that are offset from the Serpens centroid by $\sim 2.5$ mas, or 5 \kms\ (see Figure~\ref{fig:pmdunham}).  Three of the four stars W40 and Serpens South with VLBI astrometry also have proper motions consistent with this offset in proper motion.  In contrast, Figure~\ref{fig:pm_serpsouth} shows that a cluster of stars (brighter than $G<19$) have proper motions consistent with the wider Serpens region.  No other overdensity is detected at the visual location of Serpens South and W40 (for stars with the offset proper motion) or in proper motion (for stars located in W40 and Serpens South).

One interpretation is that Serpens South and W40 have projected velocities on the sky that are offset by 5 \kms\ from the other clusters within Serpens.  Most of the candidate members from the \citealt{povich13} and \citet{dunham15} catalogs are optically faint ($G>19$ and are excluded our broader overview.  The candidate members with proper motions consistent with the larger Serpens complex would then be distributed members, located in the foreground of the dark cloud rather than within the active star-forming clouds.
  An alternate explanation is that the candidate members with offset proper motions in Figure~\ref{fig:pmdunham} are field stars, since they overlap with the distribution of proper motions for stars from 300--500 pc over the full downloaded area.   The offset of $\sim 5$ \kms\ on the sky is larger than expected based on the similarity of line-of-sight velocities to cold gas in the W40 and Serpens South molecular clouds \citep{nakamura11,shimoikura15}, compared to the velocities of Serpens Main and Serpens far-South \citep{dame85,davis99}.   
 However, some expansion is seen in the gas in W40, related to the \ion{H}{2} bubble \citep{zhu06,pirogov13,shimoikura15}, and may contribute to a larger dispersion in proper motions for newly-formed stars. 
  The proper motions of the faintest objects may also suffer from larger systematic errors than expected.  Distinguishing between these explanations will require spectroscopy of candidate members to evaluate youth.

\subsubsection{Serpens Northeast}

The stellar population of Serpens Northeast is located at 478 pc, consistent with the dust extinction distance of $460\pm35$ pc.  This subcluster is the most distance cluster of young stars within the Serpens star-forming cloud complex, as defined in this paper.  Much of the optically-thick dust in the Aquila Rift is also located to the north and east of Serpens Main.  This distance may indicate some connection between Serpens Star-Forming Regions and the Aquila Rift, since the clouds in the Aquila Rift nearby to Serpens Northeast are located at $\sim 600$ pc.  Serpens Northeast includes some stars $\sim 2$ deg (16 pc) northeast of the main optical cluster, located along the direction of the Aquila Rift dust cloud. 

The optical appearance of Serpens West is visually split into two by an optically-thick dust cloud, like a dust sandwich with the optically-bright stars around the dust.  Some protostars and disks are present within the cloud \citep{dunham15}.   The location of stars on the color-magnitude diagram as well as the (candidate) protostar/disk ratio, which suggests that Serpens Northeast is older than Serpens Main but younger than Serpens far-South.
The populations north and south of the cloud are indistinguishable in distance.

The projected morphology of Serpens Northeast, with an optically-bright population around the dust, may be explained if the stars were either formed in a centralized factory and flung outward, or if the stars were born near their current location and have been eroding the nearby cloud. The stars in the northern half of the sandwich have proper motions of $+0.15$ mas/yr, relative to the proper motions of southern stars (Table~\ref{tab:dustwich}).  However, if both populations had formed at the same spatial location, they would have had to have been flung out $\sim 20$ Myr ago, older than the age estimated from the color-magnitude diagram.  Even more challenging, the standard deviation in the proper motions is $0.4$ mas/yr, so many of the stars in the north are traveling south, and vice versa, at a velocity consistent with the velocity dispersion measured for protostars \citep[e.g.][]{Kirk10,heiderman15}.   Populations to the east and west also do not correspond to a past origin in some centralized location within the present-day cloud.  

Some of these stars have either come from the opposite direction of the current location of the dark cloud, or their proper motions have already been altered by internal dynamics within the small cluster.  The optically-bright stars have likely formed out of what is now an eroded cloud, while a few protostars are still forming where the cloud is still optically-thick.  This erosion does not conflict with the possibility that stars are also being flung out from the birth filament at the lower velocities that have been observed in the Orion Nebula \citep{mairs16,stutz16}.

\begin{table*}[!t]
\begin{center}
\caption{Proper Motions of Stars in Serpens Northeast$^a$}
\label{tab:dustwich}
\begin{tabular}{ccccccccc}
\hline
N & RA Range & Dec Range & PM(RA) & $\sigma(PM-RA)$  & PM(Dec) & $\sigma(PM-RA)$ & $d$ & $\sigma(d)$\\
\hline
21 & 279.2 to 279.6 & -0.6 to -0.1 & 2.69 & 0.35 & -8.18 & 0.40 & 472 & 15\\
30 & 279.2 to 279.6 & 0.2 to 0.8 &  2.51 & 0.38 & -8.05 & 0.39 & 489 & 20\\
20 & 279.65 to 280.1 & 0.0 to 1.0 & 2.42 & 0.40 & -8.21 & 0.31 & 484 & 19\\
12 & 278.8 to 279.2 & 0.0 to 1.0 &  2.93 & 0.38 & -8.03 & 0.81 & 482 & 21\\
\hline
\multicolumn{9}{l}{$^a$For stars with $\varpi/\sigma(\varpi)>10$, from 440--530 pc, and within 1.7 mas/yr of the proper motion centroid.}
\end{tabular}
\end{center}
\end{table*}

\begin{figure}[!t]
\epsscale{1.2}
\plotone{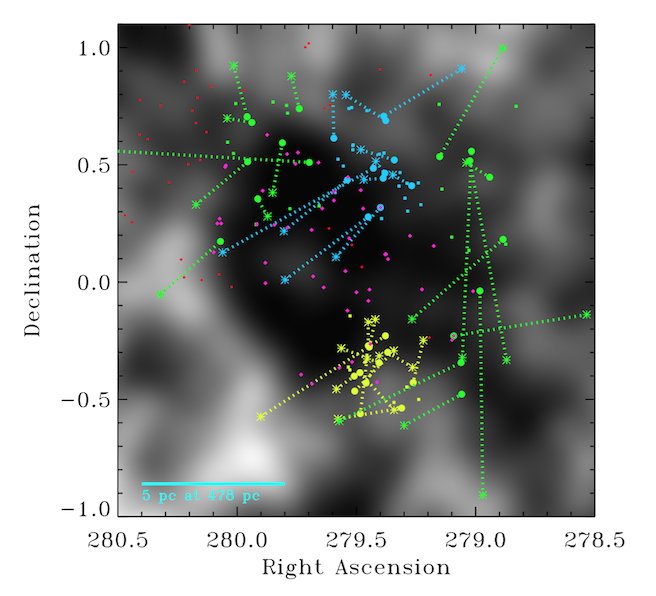}
\caption{Proper motions of stars in Serpens Northeast, showing current positions (circles) and positions 2 Myr ago (asterisks) for stars with $\varpi/\sigma(\varpi)>20$.  The colored squares have $10<\varpi/\sigma(\varpi)<20$ and are included in the analysis, but with past positions not shown for ease of visualization.  Purple diamonds show protostars and disks from the \citep{dunham15} catalog.  The blue and yellow colors show specific overdensities to the north and south of the cloud, with proper motions inconsistent with an origin within the present-day cloud.  The green outer regions are consistent with this interpretation.  Red circles shown here are likely members that are excluded because they are outside the current overdensities. The average proper motion of Serpens Northeast has been subtracted from all stellar proper motions shown here.}
\label{fig:pm_serpne}
\end{figure}

\subsubsection{Serpens far-South}
This cluster pokes out of the extreme southern end of the dark cloud, directly south of the Serpens Main and Serpens South clusters.  The optical cluster is located $\sim 1^\circ$ southeast of the optically-thick dust in Serpens far-South, where some protostars are present \citep{rumble15}.  This region is not well studied.  The {\it Spitzer} observations of Serpens \citep[e.g.][]{dunham15} were defined by the dust maps and therefore missed the optical counterparts.
Sh 2-62, powered by the Herbig Be star MWC 297, is located on the eastern end of the cluster.
The \ion{H}{2} region Sh 2-61, powered by the Herbig Be star HBC 284, is projected on the southwest edge of the cluster but has {\it Gaia} astrometry consistent with a background object.    
Many of candidate young stars in the \citet{dunham15} catalog are found to be non-members.  Based on the color-magnitude diagrams, this region is the oldest of the active star-forming complexes in Serpens.

An overdensity of stars extends 2.8 deg (19 pc) to the southwest of the main optical cluster, ending in a group of $\sim 15$ stars located at $\alpha=276$ deg, $\delta=-6$ to $-7$ deg.  The proper motion of this small group of PM(RA)=1.7 mas/yr and PM(Dec)=$-8.3$ mas/yr is offset by 0.7 mas/yr from the main cluster, so this may be a distinct group from Serpens far-South.  The location of these stars on the color-magnitude diagram indicate a similar age as the Serpens far-South cluster.

\subsubsection{LDN 673}

This dark cloud is located $\sim 120$ pc from the main Serpens star-forming regions.  The cloud has formed $\sim 15$ members bright enough to be identified with {\it Gaia}, with a mean distance of 407 pc and a mean proper motion similar to that of Serpens \citep[see also identification of this group by][]{rice06}.  The members of this sub-cluster, including AS 353A (with the associated HH object HH 32 discovered by \citealt{mundt83}) and V536 Aql, have similar colors and distance-corrected brightness as those in other Serpens regions.

\subsubsection{FG Aql Group}

We identify 5 stars, called here the FG Aql Group, with distances of $\sim 400$ pc and proper motions consistent with Serpens.  On the sky, this small group is located $\sim 60$ pc southeast of the ongoing star formation.
This  small group of stars is identified here only because \citet{prato08} listed them as possible young stars, following past identification and incorporation into the \citet{herbig88} catalog\footnote{Candidate members HBC 684 and FG Aql/G3 have astrometry consistent with background objects and are likely non-members, as do several other candidate young stars in this region that are listed the \citet{herbig88} catalogue.}. This group would have been detected with an analysis that combined a nearest neighbor criterion and photometry.

\subsection{The Star Formation History of the Serpens Molecular Cloud}

Within the Serpens cloud complex, about 60\% of the optically-bright stars are located within one of the three main (optical) clusters.  These past bursts of star formation, and the ongoing burst in Serpens South, are concentrated in specific spatial locations.
Some additional star formation has also occurred beyond this projected region, including in LDN 673 and the FG Aql Group.

The population found by the {\it Gaia} astrometric selection is likely older than the members that are located deep in the molecular cloud.  An age gradient is present within the full Serpens star-forming region, with Serpens South as the youngest sub-cluster and Serpens far-South as the oldest.  An age gradient is also present within each sub-cluster.   In Serpens Northeast and Serpens far-South sub-clusters, the protostellar population found by \citet{dunham15,rumble15} is still located within the cloud, while the optical population is likely older.  Such age gradients are widely observed on other star-forming regions \citep[e.g.][]{mairs16,getman18}.

The age gradients may be caused by cluster dynamics or a by change in the site of active star formation.  In the hierarchical star formation scenario of \citet{vazquez-semadeni18}, an age gradient is expected, but is most apparent for $\sim 20$ Myr old stars, because given a scatter in initial velocities, the older stars have longer time to disperse.  However, the standard deviation in proper motions of $\sim 0.5$ mas/yr ($\sim 1$ \kms) within each subclusters is inconsistent with a dynamical explanation for an age gradient; neither are the optically-bright stars the ones that happened to be flung out.  Instead, feedback from these stars \citep[e.g.][]{cunningham18} has likely eroded away the nearby molecular material over several Myr.

A cluster located at 430 pc with a velocity dispersion of $\sim 1$ \kms, with no internal dynamics, would expand by $\sim 1$ deg every 10
Myr.  Over $\sim 50$ Myr, any clustered star formation would likely remain in a cluster
and would likely not move far from the birth site, unless the velocity dispersion were significantly larger than those observed here.  
Two older clusters, Collinder 359 and IC 4665, were found here with {\it Gaia} DR2 astrometry \citep[see also][]{cantat18} and may be related to a burst of star formation tens of Myr ago.   Collinder 359, located at a distance of 551 pc, is visually separated by 85 pc on the sky from the northern clusters of Serpens.  Given the relative proper motions \citep{cantat18}, the cluster would have overlapped with Serpens Northeast (in projected space) 16 Myr ago, which is likely younger than the cluster age, estimated to be $\sim60$ Myr by \citet{lodieu06}.  IC 4665, located at 350 pc, is a less promising candidate.  The cluster, located at a distance of $350$ pc, has a proper motion that would have also placed it visually northeast of the Serpens clouds 13 Myr ago, much younger than the estimated age of 40 Myr by  \citep{dias02}.  Any visual association between IC 4665 and Serpens is likely coincidental, while the relationship between Collinder 359 and Serpens is unclear.

The lack of clusters and young stars beyond this immediate region indicate that the star formation in Serpens is a recent burst.  Because our selection of Serpens members is independent of any photometric criterion (other than sufficient brightness for accurate astrometry in {\it Gaia} DR2), we would have detected any significant star formation that occurred in the recent past.  Even if the distributed young stars are entirely from a previous epoch, the star formation rate over the past 50 Myr must have been at least $\sim 20$ times lower than the current rate.  Thus it seems likely that the recent bursts of star formation are the first generation of stars in this cloud complex.  
The possibility that a small burst 30-60 Myr ago in Serpens could have created Collinder 359 would reinforce that the star formation is very bursty.

The current epoch of star formation began within a few Myr of each other in regions that are separated by distances of 20--100 pc.  The typical sound speed in a molecular cloud of $\sim 0.2$ \kms\ leads to a crossing time of 100--500 Myr \citep[see, e.g., discussion of Upper Sco in][]{preibisch99,slesnick08}; a shock would therefore need to travel at $\sim 20$--$100$ \kms\ to connect these distant regions on a Myr timescale. 
The large separations between these groups is one of the more difficult-to-explain aspects of star-formation in this region.  This situation is not unique to the Serpens-Aquila region, as many massive star-forming regions contain several discrete clusters of stars separated by tens of parsecs \citep{2014ApJ...787..107K}. A prominent example is the NGC~6357 star-forming region, which contains three Orion-like clusters separated by $\sim$10--15~pc \citep{2014ApJS..213....1T}. NGC~6357 itself is separated from the star-forming region NGC~6334 in the same giant-molecular cloud complex by a projected distance of $\sim$60~pc.  On the other hand, the Orion Nebula Cluster and the broader Orion Region have undergone several distinct epochs of star formation \citep[e.g.][]{beccari17,kounkel18}.  This puzzle for Serpens is exacerbated, or at least now better established, by the lack of young stars from previous epochs of star formation, as would be seen over a very large area on the sky. 

Several mechanisms could potentially induce near-simultaneous star-formation at different locations in molecular clouds. For example, gravitational instabilities in a long, filamentary cloud can produce regularly spaced regions of star formation activity \citep{1996ApJ...472..673G}. Alternatively, the assembly of giant molecular clouds through a series of cloud-cloud collisions \citep[e.g.,][]{2017ApJ...850...62I} could produce multiple sites of star formation. In both Serpens Main and W40/Serpens South, kinematic studies of clouds suggest a history of cloud-cloud collisions \citep{duarte10,2017ApJ...837..154N,2018PASJ..tmp..131S}.  However, while cloud-cloud collisions may be the local trigger for a burst, it is hard to imagine any sort of cloud-cloud collision could occur at specific sites spread over such large distances, without any star-forming collisions in the recent past elsewhere along the clouds.

In principle, compression from a shockwave from a supernova or winds from hot stars could have triggered star formation in all three regions of the molecular cloud.  However, there are no indicators that a nearby supernova has occured.  The 40 Myr old cluster IC 4665 \citep{dias02} and Collinder 359 may have had a sufficient number of hot stars for winds to sweep up some gas, thereby precipitating local cloud-cloud collisions across a wide region on a timescale consistent with requirements on the travel time for a shock, though this explanation seems overly {\it ad hoc}.
In any case, given the large available gas resevoir in the Serpens-Aquila region \citep{zeilik78,andre10} and the extreme youth of the stellar population of the region, it is likely that the region is still in the early stages of its star formation history, the star formation rate is accelerating, and the observed stars are just the first to form.  

\section{Conclusions and Prospects for the Future}

We have performed an initial evaluation of the Serpens star-forming complexes and the Aquila Rift using {\it Gaia} DR2 astrometry.  The optically-bright young stars are concentrated in three clusters in the Serpens star-forming region, with distances from 383--478 pc and relative velocities between the clusters of $\sim 1$ \kms.  Some young stars are distributed throughout the region and may have formed in smaller groups.  The distances to optically-thick dust extinction of these regions are consistent with the distance to the stellar populations.  While Serpens members are identified using only astrometry, the {\it Gaia} photometry establishes that a tight selection in proper motion and distance generates a robust sample of young stars.  The populations identified here are older and more widely distributed than had been known previously,  complementing past mid-IR surveys that had identified populations within the dark molecular clouds that are not detectable with {\it Gaia}.

Some dust is present in the Serpens Cirrus, located to the southwest of the Serpens star-forming regions, has a distance of $\sim 250$ pc, consistent with past distance measurements to dust extinction in this region by \citep{straizys03}.  The Serpens Cirrus dust is not very opaque ($A_V\sim1$ mag when averaged over a large region) and is not associated with  significant star formation, although the protostar L483-mm and perhaps a few other young objects are forming in this area.  The dust in the Aquila Rift is located at $\sim 600$ pc to the northeast and at $\sim 500$ pc to the southwest of Serpens.  Since the distance of the Aquila Rift tends to increase smoothly from the southwest to the northeast, the Aquila Rift may be one huge complex that includes Serpens star-forming regions.  While it might seem surprising for the cloud to extend 300 pc along our line-of-sight, the Aquila Rift spans at least 250 pc across our full field of view, when projected on the sky.

The active star formation in Serpens over the past few Myr has occured across a distance of $\sim 100$ pc, in distinct sites that are typically $\sim 5$ pc across.  The absence of an older population indicates that little star formation occurred prior to this current activity.  
This strongly suggests some sort of external trigger traveling at 20-100 \kms\ to generate the star formation.  If the ages are older than inferred here, then the speed of the trigger could be reduced but would not alter the general picture.  

In Serpens Northeast, several protostars are located within the optically thick cloud, while optically bright stars are located north and south of the cloud, like a dust sandwich.  The proper motions of these optically-bright stars indicates that they were likely formed near their present locations and have eroded  way their parent molecular cloud.  Their positions and proper motions are inconsistent with having been formed in a central star formation factory and then flung out to their current location.

This evaluation should be treated as an initial analysis of the {\it Gaia} view of the Serpens-Aquila region.  Future data releases will improve astrometric precision for optically fainter sources, which will allow us to increase sample sizes and thereby better probe the low-mass and embedded populations.  The analysis of line-of-sight dust extinction should improve by adding multiple regions of dust and by incorporating the brightness and colors of objects, rather than just star counts.  A full census of the population should assign membership probabilities to each star.  Analyses of individual stars would improve age and mass estimates.

Despite these caveats to our work, we are still able to identify a new, optically bright population in Serpens  that was not previously known, as well as resolve nagging doubts about the distances to W40/Serpens South and other regions in the Aquila Rift and Serpens Cirrus.  This work was performed almost {\it exclusively} with {\it Gaia} data.

\section{Acknowledgements}

We request that any citation to this paper also cite the primary papers that describe {\it Gaia} DR2 data, because our own work is at best a minimal intellectual contribution to understanding this region with {\it Gaia}.  

We thank the anonymous referee for clear, careful, and thoughtful comments, which have improved the quality of the paper.  GJH and MK thank Lynne Hillenbrand for discussions about Gaia and on the draft, including detailed suggestions on figures and captions. GJH also thanks Kaitlin Kratter for a discussion of star formation theory, Logan Francis for an initial discussion of extinction maps in comparison to {\it Gaia} star count maps, Neal Evans for a discussion of L483, and Martin Smith for many discussions about {\it Gaia}.  

This work has made use of data from the European Space Agency (ESA) mission
{\it Gaia} (\url{https://www.cosmos.esa.int/gaia}), processed by the {\it Gaia}
Data Processing and Analysis Consortium (DPAC, \url{https://www.cosmos.esa.int/web/gaia/dpac/consortium}). Funding for the DPAC has been provided by national institutions, in particular the institutions
participating in the {\it Gaia} Multilateral Agreement.  This research has made use of the Spanish Virtual Observatory (http://svo.cab.inta-csic.es) supported from the Spanish MINECO/FEDER through grant AyA2014-55216

GJH is supported by general grants 11773002 and 11473005 awarded by the National
Science Foundation of China.  CFM is supported by an ESO Fellowship.  DJ is supported by the National Research Council Canada and by an NSERC Discovery Grant.  AB acknowledges the research grant $\#11850410434$ awarded by the National Natural Science Foundation of China through a Research Fund for International Young Scientists
and China post-doctoral General grant.

\begin{figure*}[!t]
\includegraphics[width=0.33\textwidth, trim=50 370 50 45,clip]{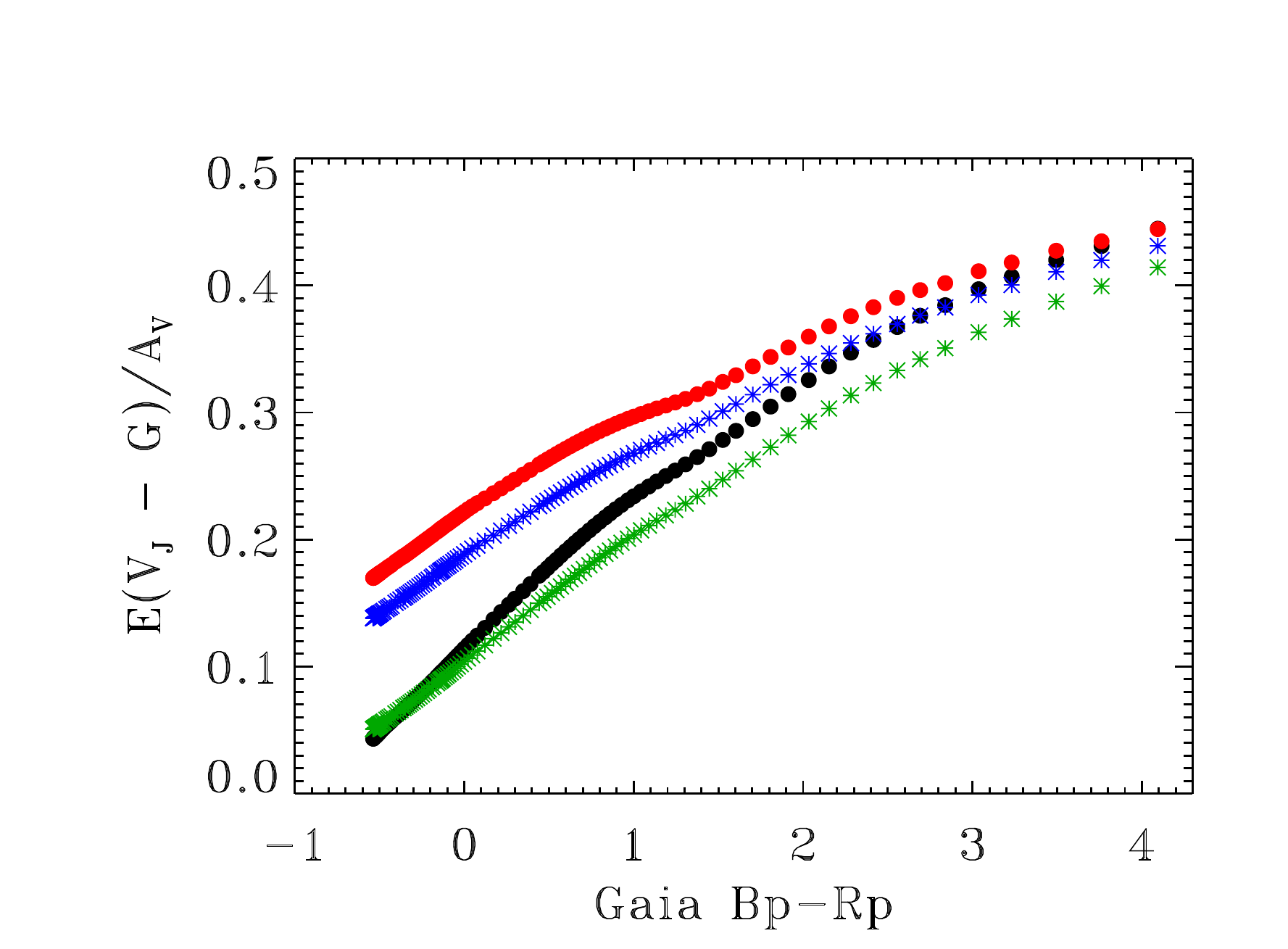}
\includegraphics[width=0.33\textwidth, trim=50 370 50 45,clip]{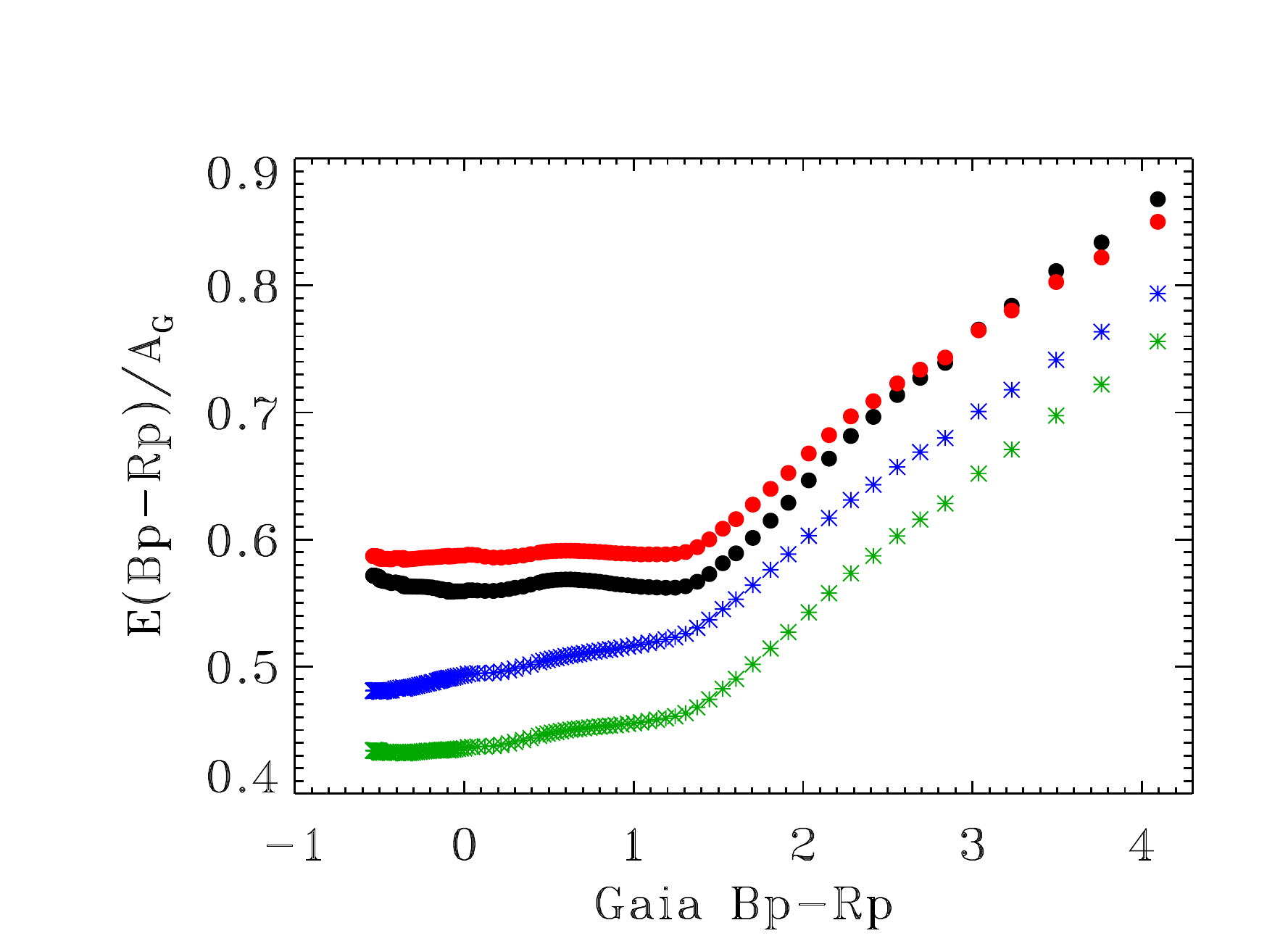}
\includegraphics[width=0.33\textwidth, trim=50 370 50 45,clip]{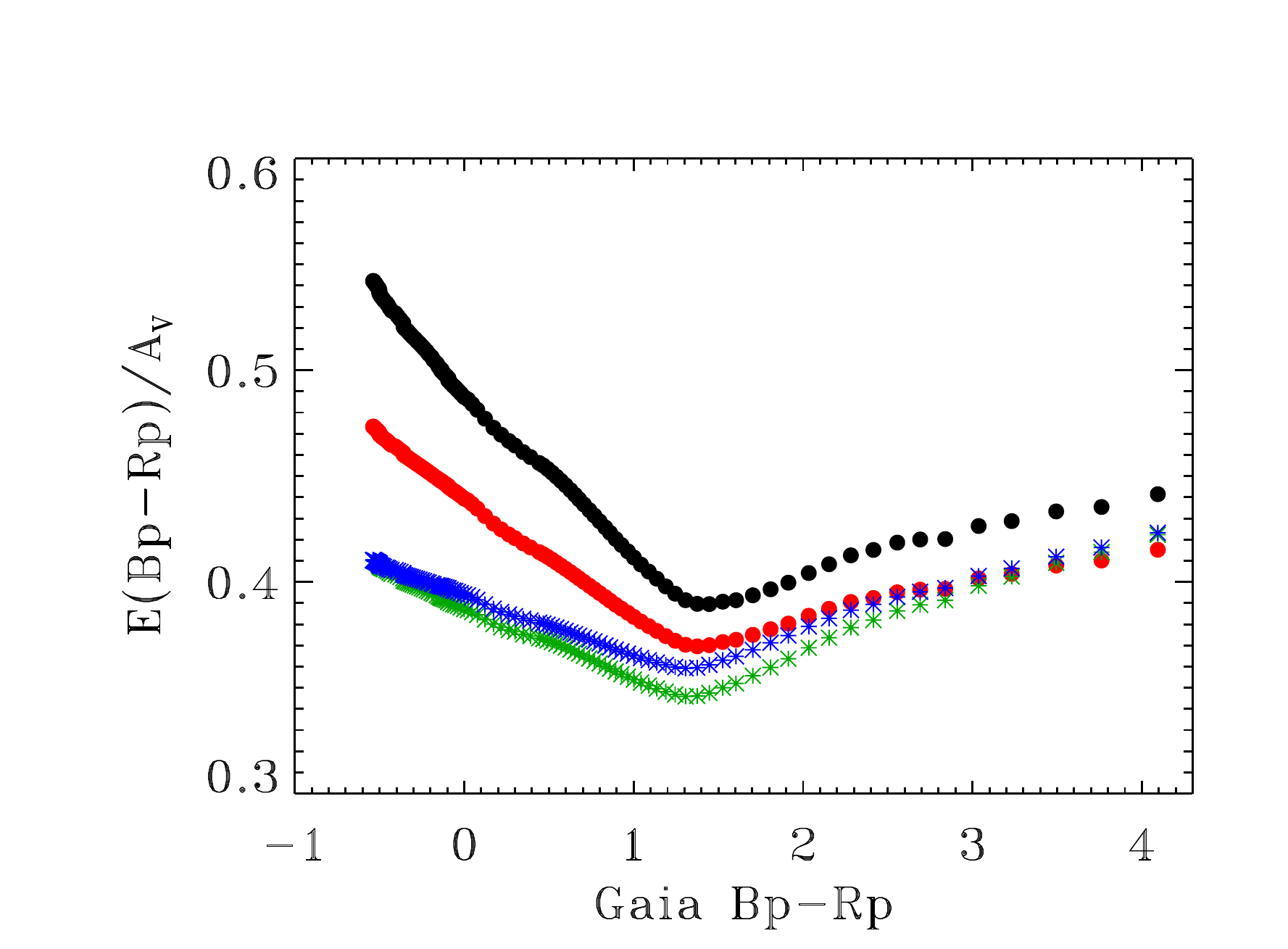}
\caption{Extinction vectors for Gaia $A_G/A_V$ (left), $E(B_P-R_P)/A_G$ (middle), and $E(B_P-R_P)/A_V$ (right) versus stellar color $B_P-R_P$, for $A_V=1$ and $5$ mag for $R_V=3.1$ (red and black circles, respectively) and $A_V=1$ and $5$ mag for $R_V=5.5$ (blue and green asterisks, respectively).}
\label{fig:extvectors}
\end{figure*}

\begin{table*}
\begin{center}
\caption{Color changes for $A_V=1$ mag$^a$}
\label{tab:extvectors}
 \begin{tabular}{rccccccc}
&& \multicolumn{3}{c}{$R_V=3.1$} & \multicolumn{3}{c}{$R_V=5.5$} \\
 T (K) & $B_P-R_P$ &  $E(V-G)$ & $E(V-B_P)$ & $E(B_P-R_P)$ & $E(V-G)$ & $E(V-B_P)$ & $E(B_P-R_P)$ \\
 \hline
 2600 &$    4.09$&$    0.44$&$    0.03$&$    0.44$&$    0.41$&$    0.02$&$    0.42$\\
 2800 &$    3.49$&$    0.42$&$    0.02$&$    0.43$&$    0.39$&$    0.01$&$    0.41$\\
 3000 &$    3.04$&$    0.40$&$    0.02$&$    0.43$&$    0.36$&$    0.01$&$    0.40$\\
 3200 &$    2.69$&$    0.38$&$    0.01$&$    0.42$&$    0.34$&$    0.01$&$    0.39$\\
 3400 &$    2.42$&$    0.36$&$    0.01$&$    0.42$&$    0.32$&$    0.01$&$    0.38$\\
 3600 &$    2.15$&$    0.34$&$    0.01$&$    0.41$&$    0.30$&$    0.00$&$    0.37$\\
 3800 &$    1.91$&$    0.31$&$    0.00$&$    0.40$&$    0.28$&$    0.00$&$    0.36$\\
 4000 &$    1.70$&$    0.29$&$    0.00$&$    0.39$&$    0.26$&$    0.00$&$    0.36$\\
 4200 &$    1.53$&$    0.28$&$   -0.00$&$    0.39$&$    0.25$&$   -0.00$&$    0.35$\\
 4400 &$    1.38$&$    0.26$&$   -0.01$&$    0.39$&$    0.23$&$   -0.00$&$    0.35$\\
 4500 &$    1.31$&$    0.26$&$   -0.01$&$    0.39$&$    0.23$&$   -0.00$&$    0.35$\\
 5000 &$    1.04$&$    0.24$&$   -0.02$&$    0.41$&$    0.21$&$   -0.01$&$    0.35$\\
 5500 &$    0.86$&$    0.22$&$   -0.04$&$    0.42$&$    0.19$&$   -0.02$&$    0.36$\\
 6000 &$    0.71$&$    0.20$&$   -0.05$&$    0.44$&$    0.18$&$   -0.03$&$    0.37$\\
 6500 &$    0.57$&$    0.19$&$   -0.06$&$    0.45$&$    0.16$&$   -0.04$&$    0.37$\\
 7000 &$    0.44$&$    0.17$&$   -0.07$&$    0.46$&$    0.15$&$   -0.04$&$    0.37$\\
 8000 &$    0.22$&$    0.14$&$   -0.08$&$    0.47$&$    0.13$&$   -0.05$&$    0.38$\\
 9000 &$    0.02$&$    0.12$&$   -0.10$&$    0.49$&$    0.11$&$   -0.06$&$    0.39$\\
10000 &$   -0.07$&$    0.10$&$   -0.11$&$    0.49$&$    0.10$&$   -0.06$&$    0.39$\\
25000 &$   -0.43$&$    0.06$&$   -0.15$&$    0.53$&$    0.06$&$   -0.09$&$    0.40$\\
33000 &$   -0.50$&$    0.05$&$   -0.16$&$    0.54$&$    0.05$&$   -0.09$&$    0.41$\\
41000 &$   -0.51$&$    0.05$&$   -0.16$&$    0.54$&$    0.05$&$   -0.09$&$    0.41$\\
49000 &$   -0.52$&$    0.05$&$   -0.16$&$    0.54$&$    0.05$&$   -0.09$&$    0.41$\\
57000 &$   -0.53$&$    0.04$&$   -0.16$&$    0.54$&$    0.05$&$   -0.09$&$    0.41$\\
65000 &$   -0.54$&$    0.04$&$   -0.16$&$    0.54$&$    0.05$&$   -0.10$&$    0.41$\\

\hline
\multicolumn{8}{l}{Calculated from synthetic spectra, as described in Appendix A.}
\end{tabular}
\end{center}
\end{table*}

\section{Appendix A:  Extinction Vectors for Gaia photometry}

The {\it Gaia} photometric bands are broad enough to introduce large differences in the color change caused by extinction as a function of the instrinsic stellar color.  We calculate extinction vectors by applying extinction curves of \citet{fitzpatrick99} to synthetic stellar spectra from the Phoenix/NEXTGEN grid (BT-Settl version from \citealt{allard12} with solar abundances from \citealt{anders89} and $\log g=4$), and subsequently convolving the reddened spectrum in photon units with the filter transmission curve.  Filter curves for {\it Gaia} DR2 and the spectral templates were obtained from the Spanish Virtual Observatory.

Figure~\ref{fig:extvectors} shows the change in {\it Gaia} $G$, $B_P$, and $R_P$ magnitudes for extinctions of $A_V=1$ and $A_V=5$ mag and total-to-selective extinction coefficients of $R_V$=3.1 (adopted in this paper) and 5.5 (see also Table~\ref{tab:extvectors}).  The change in $G$ versus $A_V$ is less for redder stars because most of the $G$-band photons are red and suffer less from extinction.
The change in $B_P-R_P/G$ shows a sharp increases versus extinction $A_G$ for stars cooler than $B_P-R_P=1.5$, because the $R$-band photons also dominate the $G$-band photometry.  The $E(B_P-R_P)$/$A_V$ curve varies much less (within 0.1 mag) for stars with intrinsic colors between $B_P-R_P$ between $0$ and $4$.  For low-mass pre-main sequence stars, the effect of extinction on $B_P-R_P$ versus $G$ is nearly parallel to isochrones, as is also the case with $V-I$ versus $V$ \citep[e.g.][]{sicilia05}.

\section{Appendix B:  The distance and extinction to a sheet of dust}

\begin{figure}[!t]
\includegraphics[width=0.49\textwidth, trim=40 300 40 40,clip]{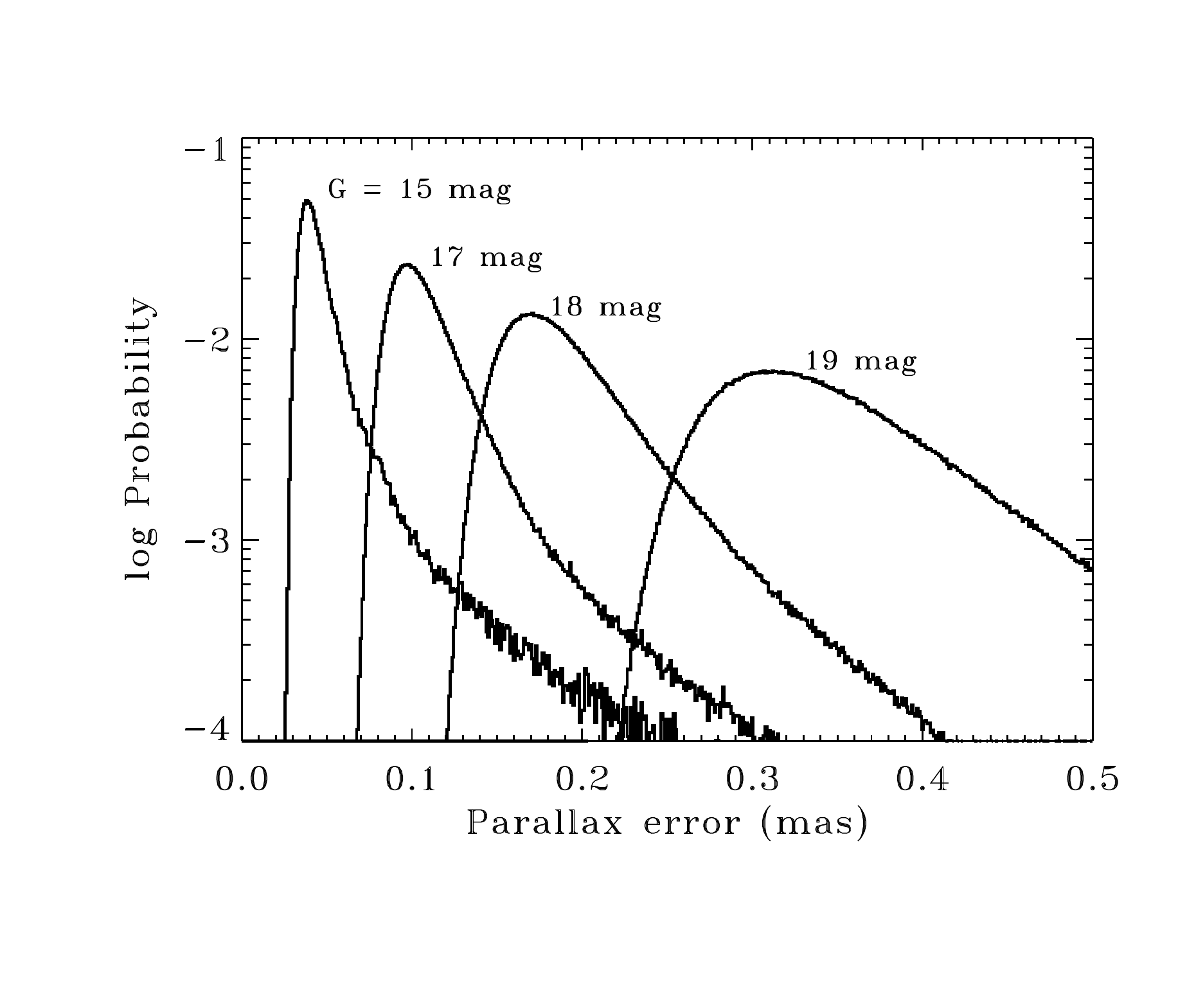}
\vspace{-15mm}
\caption{The probability distribution of errors for the 140 million stars in the downloaded field, shown here for brightness bands within 0.2 mag of $G=15$, $17$, $18$, and $19$ mag, in bins of 0.001 mas/yr.  In our simulations, the errors are randomly assigned for the updated brightness with an empirical probability distribution, since the distributions are not well described by Gaussian profiles.}
\label{fig:gaiaerrors}
\end{figure}

In this appendix, we describe our code to measure the distance to dust
extinction.  Dark clouds extinguish emission from background stars,
thereby making them fainter and redder.  We assume that the cloud is a thin sheet of dust at a single distance, and subsequently simulate stellar populations to estimate the distance $d$ and extinction $A_G$ of the sheet. This idea is similar to distance measurements to dark clouds using {\it Gaia} DR2 by \citet{yan19} and \citet{zucker19}, {\it Gaia} DR1 by \citet{voirin18}, and {\it Hipparcos} by \citet{whittet97}, among others, with differences in implementation.

Each simulation includes 500,000 stars that are randomly selected from a parent (or off-cloud) population.  If a star is in front of the dust, then the measured stellar parallax, parallax error, and brightness are adopted. If a star is behind the dust, then the  {\it Gaia} $G$-band photometry is made fainter by $A_G$.  A new parallax error for this fainter star is adopted by randomly selecting the parallax error for the $G$-band photometry, based on the distribution of parallax errors in our full dataset (Figure~\ref{fig:gaiaerrors}).  All parallaxes, for stars in front of and behind the cloud\footnote{Stars with negative parallaxes in the parent sample are assumed to have a parallax equal to 0 mas.  This choice does not affect our final results, since the clouds are located within 1 kpc.}, are then adjusted by adding the parallax error multipled by a number randomly selected from a Gaussian distribution.

The parent population is the population that would be present along the line of sight, if extinction were not present.  Within the region downloaded for this paper, 28 million stars have $G<19$ and are located in regions that are not heavily extincted.  The stellar density depends on the distance from the galactic plane.  More stars are located south of the plane, likely because of modest extinction along and just north of the plane.  Simulations are therefore run for three different parent populations:  stars with the same galactic latitude, stars with the opposite galactic latitude, and stars within some annulus around the region.

The simulated distribution is then compared to the observed distribution of stars in the line of sight using a two-sided Kolmogorov-Smirnov test.  The distribution is scaled to the number of observed stars.  Since the clouds in the Serpens Star-Forming Region are located at $\sim 400-500$ pc, the comparison is restricted to stars with distances (or simulated distances) less than 1000 pc that are measured (or simulated to be measured) to better than 5\% and are brighter (or simulated to be brighter) than $G=19$ mag.  Separate simulations are run for sources with parallaxes measured to better than 20\%.  The comparison avoids stars with proper motions within 2 mas/yr of the Serpens clusters.  Sources with excess astrometric noise of greater than 2 mas are excluded from the parent sample.

Figure~\ref{fig:contourdist} shows results for six different regions.  In the simulations, the dropoff in stars versus distance is somewhat gradual, with a cloud distance that is roughly in the center of the dropoff.  The width of this dropoff increases as the  average $\varpi/\sigma(\varpi)$ of a sample decreases.
Even for optically thick clouds, the dropoff can occur at distances closer than expected:  stars that are closer than the cloud have errors that can make them appear at larger distances, but there are no stars at larger distances that may appear closer than the cloud.  Clouds are often patchy, with a range of extinctions; multiple clouds within some region may also be spread across a range of distances.    Moreover, the use of samples with  ($\varpi/\sigma(\varpi)>20$ and $>5$ produce different extinctions because they are sensitive to different types of stars.  The extinctions measured here are therefore not reliable.  Distances obtained from different samples and parallax care always consistent.

This simulation assumes that the errors are accurately evaluated.  The parallax error is assumed to be independent of object distance and object color.  The excess astrometric noise is also assumed to be independent of distance and color.  The parent distribution of {\it Gaia} sources is assumed to be independent of crowding.  
The comparison between the simulated and observed distribution is restricted to the distribution of distances in the two samples, so the total number of stars is essentially scaled to match.  In principle, this method could be adapted to also test the total stellar population and additional parameters, such as the distribution of brightnesses and colors.

These calculations assume that the extinction occurs in a sheet located at a single distance and with a single extinction.  However, molecular clouds are clumpy and may be distributed over some distance.  For instance, the Serpens Main cloud extends $\sim 1$ deg, or 7 pc, from south to north, and likely has a similar radial distance.  Multiple clouds may be located along the same line of sight, or different lines of sight but within the large beam.

\begin{figure}[!t]
\includegraphics[width=0.49\textwidth, trim=40 300 40 40,clip]{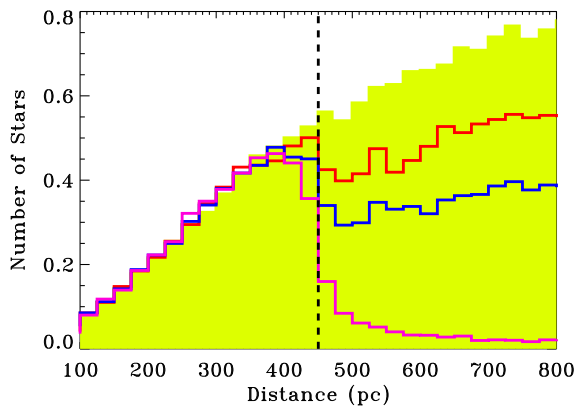}
\vspace{-15mm}
\caption{Examples of simulated stellar distributions with $\varpi/\sigma(\varpi)>20$ versus distance for an off-cloud population (yellow shaded region), at $A_G=1$, $2$, and $6$ mag (red, blue, and purple lines, respectively), with star counts binned to 25 pc.}
\label{fig:example}
\end{figure}

\section{Appendix C:  A Description of the Online Source List}

 In an online electronic catalog (see Table~\ref{tab:starprops}), we provide a list of all possible Serpens members within 7$^\circ$ of ($278,-2$), within 3 mas/yr of a proper motion of ($2.5,-8.4$) mas/yr, located at a distance of 320--550 pc, with $\varpi/\sigma(\varpi)>5$ and with an excess astrometric noise of $<2$ mas.   Beyond the 7$^\circ$ radius, the table also includes several objects near LDN 673, including AS 353A, and the FG Aql group.

The table identifies the highly-likely members by listing the region membership, the same target that are selected for the HR diagrams in Figure \ref{fig:hrdiags2}.  Similarly, the likely members of the distributed Serpens star-formation are listed as ``Distributed''.  The full online table is liberal in including many non-members, identified by membership listed as ``N/A''.

The table also includes nearest neighbor calculations as a way of identifying likely members and small, isolated groups of stars.  For this calculation, a parent dataset is established of all stars between 350--520 pc with $\varpi/\sigma(\varpi)>20$ and within 3 mas/yr of the Serpens proper motion centroid.  For each star in the full sample, we then calculate the 7th nearest spatial neighbor within the parent dataset.  For each star, we also calculate the 7th nearest proper motion neighbor for stars within a $1^\circ$ radius within the parent dataset.  The Serpens members especially stand out as having low 7th nearest neighbors, with typical distances of $<0.4^\circ$ compared to an average distance of $\sim 0.7^\circ$ for field stars.  Most members also have 7th nearest neighbor in proper motion of $<0.5$ mas/yr, while field stars have an average of value of 1.3 mas/yr.  This nearest neighbor analysis alone would produce modest contamination rates, especially away from the main region.  A nearest neighbor criterion combined with a photometric selection would have readily identified the LDN 673 group and the FG Aql group as likely members, along with the far soutwest extent of Serpens far-South.  In addition, the two stars 2MASS J18435376+0320524 and 2MASS J18430797+0321056 are likely Serpens members and may form a small group.

\bibliographystyle{apj}
\bibliography{main}

\begin{table*}
\begin{center}
\caption{Evaluating Past Membership Surveys}
\label{tab:members}
\begin{tabular}{lcccccc}
&&& \multicolumn{2}{c}{Non-member} &\multicolumn{2}{c}{Likely member} \\
\hline
& & & & $\varpi>2\sigma$ & no $\varpi$ & $\varpi<2\sigma$ \\
~~~~~Catalog &  Total & Matches & PM$>4\sigma^a$ & PM$<4\sigma^a$,
 & PM$<4\sigma^a$ & PM$<4\sigma^a$ \\
\hline
\multicolumn{7}{c}{Serpens Main}\\
\hline
Total Serpens Main & 494 & 180 & 65 & 7 & 24 & 84 \\
\hline
Dunham Total$^b$ & 227 & 104 & 20 & 5 & 16 & 63\\
~~~Class 0/I, Flat &  52 & 3 & 1 & 0 & 0 & 2\\
~~~Disks     & 131 & 73 & 6 & 4 & 15 & 48 \\
~~~Evolved   &  39 & 28 & 13 & 1 & 1 & 13\\
~~~(Likely AGB) & 5 & 5 & 5 & 0 & 0 & 0\\
\hline
\citet{giardino07} & 56 & 23 & 2 & 0 & 6 & 15 \\
\citet{eiroa08} & 253 & 74 & 33 & 2 & 14 & 25 \\
\citet{winston10} & 66 & 47 & 12 & 0 & 11 & 24 \\
\citet{gorlova10} & 19 & 3 & 0 & 0 & 1 & 2\\
\citet{erickson15} & 63 & 54 & 17 & 1 & 1 & 35\\
\citet{getman17} & 159 & 40 & 10 & 1 & 7 & 22\\
\hline
\hline
\multicolumn{7}{c}{Serpens South and W40$^c$}\\
\hline
\hline
Dunham Total$^{b}$ & 725 & 153 & 126 & 5 & 18 & 4 \\
~~~Protostars & 117 & 2 & 2 & 0 & 0 & 0\\
~~~Disks & 257 & 45 & 25 & 2 & 16 & 2\\
~~~Evolved & 351 & 106 & 99 & 3 & 2 & 2\\
~~~(Likely AGB) & 69 & 37 & 35 & 1 & 1 & 0\\
\hline
\citet{getman17} & 645 & 18 & 12 & 0 & 4 & 2\\
\citet{winston18} & 66 & 16 & 10 & 0 & 3 & 3 \\
\hline
Povich Disks+Protostars & 751 & 31 & 23 & 2 & 0 & 6\\
~~~Protostars & 281 & 5 & 2 & 2 & 0 & 1\\
~~~Disks & 470 & 26 & 21 & 0 & 0 & 5 \\
~~~Uncertain & 244 & 8 & 6 & 0 & 0 & 2\\
\hline
\hline
\multicolumn{7}{c}{Serpens Northeast$^c$}\\
\hline
Dunham Total$^{b}$ & 224 & 97 & 77  & 9 & 6 & 5 \\
~~~Protostars & 16 & 3 & 2 & 1 & 0 & 0\\
~~~Disks & 36 & 14 & 3 & 5 & 3 & 3\\
~~~Evolved & 172 & 80 & 72 & 3 & 3 & 2\\
~~~(Likely AGB) & 41 & 32 & 32 & 0 & 0 & 0\\
\hline
\hline
\multicolumn{7}{c}{Serpens far-South$^c$}\\
\hline
Dunham Total$^b$ & 265 & 112 & 99 & 1 & 5 & 7\\
~~~Class 0/I, Flat  & 10  &  0 & 0 & 0 & 0 & 0\\
~~~Disks & 34 & 14 & 6 & 0 & 3 & 5\\
~~~Evolved &  221 & 98 & 93 & 11 & 2 & 2\\
~~~(Likely AGB) & 45 & 40 & 39 & 0 & 0 & 1\\
\hline
\hline
\multicolumn{7}{c}{Spatially Distributed Serpens$^c$}\\
\hline
Dunham Total$^b$ & 105 & 59 & 58 & 1 & 0 & 0\\
~~~Class 0/I, Flat & 4 & 1 & 1 & 0 & 0 & 0\\
~~~Disks &  3 & 2 & 1 & 1 & 0 & 0\\
~~~Evolved & 97 & 56 & 56 & 0 & 0 & 0\\
~~~(Likely AGB) & 55 &  52 & 52 & 0 & 0 & 0 \\
\hline
\multicolumn{7}{l}{$^a$Here $4-\sigma$ in proper motion as $\sim2$ mas/yr.}\\
\multicolumn{7}{l}{~~~The distance criterion of $2\sigma$ is based on parallax uncertainty.}\\
\multicolumn{7}{l}{$^b$Likely AGB stars are excluded from all totaled numbers.}\\
\multicolumn{7}{l}{$^c$Stars labeled "Aquila" by \citet{dunham15} are allocated to different regions as follows:}\\
\multicolumn{7}{l}{~~~Serpens Northeast:  Declination $>-1^\circ$}\\
\multicolumn{7}{l}{~~~Serpens South+W40:  Declination $>-1^\circ$}\\
\multicolumn{7}{l}{~~~Serpens far-South:  within 1.5$^\circ$ of (}\\
\multicolumn{7}{l}{~~~Serpens Distributed:  Everything else labeled "Aquila"}\\
\end{tabular}
\end{center}
\end{table*}

\begin{table*}[!b]
\label{tab:starprops}
\begin{center}
\caption{High Probability Candidate Members of the Serpens Star-Forming Region$^a$}
\begin{tabular}{ccccccccccccc}
RA & Dec & PM(RA) & PM(Dec) & $\varpi$ & $\sigma(\varpi)$ & $G$ & $Bp$ & $R_p$ & astr.err. & $N_7$ & $N{\rm PM}_7$ & Region\\
& & mas/yr & mas/yr & mas & mas & &&& mas & $^\circ$ & mas/yr & \\
\hline
    290.06057  & $     11.25336$ & $   3.406$ & $ -10.253$ & $   2.484$ & $  0.064  $ &    15.501   &     16.925 &     14.298 &  0.25  & 0.35  & 0.86 &    LDN 673 \\
279.73961  & $      0.73944$ & $   2.643$ & $  -8.491$ & $   2.149$ & $  0.047  $ &    12.928   &     13.814 &     11.996 &  0.12  & 0.10  & 0.16 &         NE \\
    278.88497  & $      0.18238$ & $   2.015$ & $  -7.629$ & $   2.006$ & $  0.029  $ &    13.324   &     14.076 &     12.461 &  0.00  & 0.22  & 0.34 &         NE \\
    277.50416  & $      0.63222$ & $   3.365$ & $  -7.908$ & $   2.159$ & $  0.070  $ &    13.448   &     14.709 &     12.326 &  0.40  & 0.09  & 0.28 &       Main \\
    279.06096  & $     -0.34338$ & $   1.783$ & $  -7.794$ & $   1.967$ & $  0.030  $ &    13.545   &     14.506 &     12.558 &  0.13  & 0.24  & 0.29 &         NE \\
    279.95607  & $      0.51297$ & $   2.312$ & $  -7.913$ & $   2.126$ & $  0.025  $ &    13.626   &     14.510 &     12.697 &  0.00  & 0.25  & 0.15 &         NE \\
    279.36828  & $      0.39639$ & $   3.123$ & $  -8.414$ & $   2.098$ & $  0.328  $ &    13.634   &     14.578 &     12.659 &  0.04  & 0.09  & 0.35 &         NE \\
    279.38550  & $      0.70754$ & $   3.290$ & $  -8.606$ & $   2.017$ & $  0.034  $ &    13.901   &     14.732 &     13.004 &  0.00  & 0.20  & 0.46 &         NE \\
    279.59444  & $      0.61351$ & $   2.694$ & $  -8.578$ & $   2.100$ & $  0.046  $ &    14.098   &     15.264 &     12.998 &  0.24  & 0.15  & 0.19 &         NE \\
    279.18813  & $      0.88308$ & $   3.333$ & $  -8.511$ & $   2.053$ & $  0.036  $ &    14.154   &     15.031 &     13.225 &  0.00  & 0.35  & 0.57 &         NE \\
    277.24205  & $      0.29007$ & $   2.529$ & $  -8.475$ & $   2.187$ & $  0.041  $ &    14.228   &     15.696 &     12.998 &  0.24  & 0.11  & 0.36 &       Main \\
    279.39918  & $      0.31813$ & $   2.366$ & $  -7.862$ & $   2.065$ & $  0.046  $ &    14.395   &     15.465 &     13.348 &  0.07  & 0.12  & 0.21 &         NE \\
    277.37606  & $      0.70884$ & $   2.394$ & $  -8.760$ & $   2.422$ & $  0.034  $ &    14.448   &     15.657 &     13.370 &  0.10  & 0.14  & 0.29 &       Main \\
    279.48368  & $     -0.56116$ & $   2.753$ & $  -8.662$ & $   2.098$ & $  0.036  $ &    14.563   &     15.545 &     13.578 &  0.00  & 0.17  & 0.17 &         NE \\
    277.05635  & $     -0.04708$ & $   3.301$ & $  -7.615$ & $   2.345$ & $  0.077  $ &    14.635   &     16.149 &     13.413 &  0.29  & 0.24  & 0.33 &       Main \\
    279.81014  & $      0.59363$ & $   2.628$ & $  -7.859$ & $   2.091$ & $  0.045  $ &    14.650   &     15.775 &     13.584 &  0.19  & 0.17  & 0.14 &         NE \\
    277.26643  & $      0.33920$ & $   3.425$ & $  -8.124$ & $   2.273$ & $  0.045  $ &    14.751   &     15.987 &     13.432 &  0.23  & 0.07  & 0.22 &       Main \\
    279.26180  & $     -0.42826$ & $   2.779$ & $  -8.566$ & $   2.020$ & $  0.037  $ &    14.767   &     15.807 &     13.761 &  0.09  & 0.17  & 0.14 &         NE \\
    277.53601  & $      0.97951$ & $   3.210$ & $  -8.441$ & $   2.279$ & $  0.050  $ &    14.968   &     16.173 &     13.853 &  0.21  & 0.08  & 0.21 &       Main \\
    279.45960  & $     -0.43021$ & $   2.802$ & $  -8.451$ & $   2.200$ & $  0.046  $ &    14.971   &     16.069 &     13.924 &  0.14  & 0.09  & 0.21 &         NE \\
    279.45040  & $      0.27797$ & $   2.074$ & $  -7.759$ & $   2.116$ & $  0.070  $ &    15.290   &     16.444 &     14.202 &  0.19  & 0.17  & 0.19 &         NE \\
    277.49322  & $      0.63133$ & $   2.970$ & $  -8.656$ & $   2.198$ & $  0.073  $ &    15.480   &     16.962 &     14.239 &  0.38  & 0.09  & 0.14 &       Main \\
    285.59253  & $     -5.60571$ & $   1.656$ & $ -10.015$ & $   2.443$ & $  0.042  $ &    12.783   &     13.547 &     11.894 &  0.00  & 0.47  & 1.03 &      FGAql \\
\hline
\multicolumn{9}{l}{$^a$This full table is available online.}\\
\end{tabular}
\end{center}
\end{table*}

\end{CJK*}

\end{document}